\documentclass[twocolumn,english]{revtex4}
\usepackage[T1]{fontenc}
\usepackage[latin9]{inputenc}
\usepackage{color}
\usepackage{float}
\usepackage{amsmath}
\usepackage{graphicx}
\usepackage{amssymb}

\makeatletter

\providecommand{\tabularnewline}{\\}
\newcommand{\lyxdot}{.}

\@ifundefined{textcolor}{}
{%
 \definecolor{BLACK}{gray}{0}
 \definecolor{WHITE}{gray}{1}
 \definecolor{RED}{rgb}{1,0,0}
 \definecolor{GREEN}{rgb}{0,1,0}
 \definecolor{BLUE}{rgb}{0,0,1}
 \definecolor{CYAN}{cmyk}{1,0,0,0}
 \definecolor{MAGENTA}{cmyk}{0,1,0,0}
 \definecolor{YELLOW}{cmyk}{0,0,1,0}
 }

\newcommand{\MeV}{{\rm MeV}}

\makeatother

\usepackage{babel}

\begin{document}

\title{Viscous damping of r-modes: Large amplitude saturation}

\author{Mark G. Alford, Simin Mahmoodifar and Kai Schwenzer}

\address{Department of Physics, Washington University, St. Louis, Missouri,
63130, USA}
\begin{abstract}
We analyze the viscous damping of r-mode oscillations of compact stars,
taking into account non-linear viscous effects in the large-amplitude
regime. The qualitatively different cases of hadronic stars, strange
quark stars, and hybrid stars are studied. We calculate the viscous
damping times of r-modes, obtaining numerical results and also general
approximate analytic expressions that explicitly exhibit the dependence
on the parameters that are relevant for a future spindown evolution
calculation. The strongly enhanced damping of large amplitude oscillations
leads to damping times that are considerably lower than those obtained
when the amplitude dependence of the viscosity is neglected. Consequently,
large-amplitude viscous damping competes with the gravitational instability
at all physical frequencies and could stop the r-mode growth in case
this is not done before by non-linear hydrodynamic mechanisms.
\end{abstract}
\maketitle

\section{Introduction}

A compact star is one of the most stable forms of matter in the universe.
The only instability that threatens its existence is collapse into
a black hole, triggered by unstable radial oscillation modes that
push the star below its Schwarzschild radius. However, there are other
unstable oscillation modes, the so called r-modes \cite{Papaloizou:1978zz,Lindblom:1999yk},
which damp the rotation of the star by the emission of gravitational
radiation \cite{Andersson:1997xt}. In slowly-rotating stars r-modes
themselves are damped by bulk and shear viscosity, but at high rotation
frequencies they are unstable and grow exponentially. In a companion
paper \cite{Alford:2010fd} we analyzed these instability regions
in detail. The main result of that study was that these regions vary
greatly between qualitatively different classes of stars containing
distinct phases of strongly interacting matter, but are extremely
insensitive to unknown quantitative details of the equation of state
and the transport properties within a given class. Therefore, a proper
understanding of the r-mode dynamics could in the future provide robust
signatures for the presence of exotic phases in compact stars. 

Since exponentially growing r-modes will destroy the star if their
growth is not stopped by some non-linear mechanism, the fact that
fast spinning compact stars are observed suggests that such a non-linear
damping mechanism is indeed present. More importantly, even if stopped
at a finite amplitude, r-modes still strongly emit gravitational waves
and could provide an extremely efficient mechanism for the spin-down
of compact stars \cite{Owen:1998xg} and an interesting signal for
terrestrial gravitational wave detectors. Spin-down due to r-modes
could explain the observed absence of fast-spinning young stars despite
the fact that their creation during a supernova could naturally lead
to a fast spinning remnant. For spin-down via r-modes the size of
the saturation amplitude is crucial. If the amplitude $\alpha$ is
too low, it takes too long to spin down the star; if the amplitude
is too large, $\alpha\!>\! O\left(1\right)$ , the r-mode would disrupt
the star's structure, and even before this point the r-mode could
be destroyed. If the r-mode amplitude saturates at an intermediate
value, a fast spin-down is possible. Previously, various mechanisms
for the large-amplitude behavior of r-modes have been suggested \cite{Lindblom:2000az}.
They include the coupling between different modes \cite{Bondarescu:2007jw,Arras:2002dw},
the decay into daughter modes and the eventual transformation of the
r-mode energy into differential rotation \cite{Gressman:2002zy,Lin:2004wx},
friction between different layers of the star, and surface effects
in the star\textquoteleft{}s crust \cite{Bildsten:2000ApJ...529L..33B}. 

Here we study an alternative mechanism that does not involve such
complicated non-linear dynamical or structural effects. It is present
already in a standard hydrodynamical description and exploits the
fact that at large amplitudes the damping due to bulk viscosity increases
dramatically with the amplitude \cite{Alford:2010gw,Reisenegger:2003pd,Madsen:1992sx,Shovkovy:2010xk}.
In this suprathermal regime, where the deviation from chemical equilibrium
$\mu_{\Delta}$ fulfills $\mu_{\Delta}\!\gtrsim\! T$, the viscous
damping could overcome the initial gravitational instability and saturate
the r-mode. However, as shown in \cite{Alford:2010gw}, the bulk viscosity
has a maximum as a function of the amplitude and decreases again at
even larger amplitudes. If the amplitude exceeds this critical value
then the r-mode growth cannot be stopped by viscous damping and other
non-linear dynamic effects \cite{Lindblom:2000az,Gressman:2002zy,Lin:2004wx,Arras:2002dw,Bondarescu:2007jw}
are required to saturate it. Nevertheless, we find that over a significant
region of the parameter space the suprathermal enhancement is indeed
sufficient to saturate the r-mode at a finite amplitude and the r-mode
can then efficiently spin down the star. This is in contrast to certain
non-linear hydrodynamical effects where the r-mode could completely
decay \cite{Lin:2004wx} and would not be able to cause an appreciable
spin-down of the star.

For r-modes with amplitudes sufficiently below the maximum, we give
general analytic expressions for the suprathermal damping time valid
for various forms of dense matter. For r-modes with arbitrary amplitudes,
where an analytic evaluation is not possible, we give a general expression
for the bulk viscosity damping time that includes the complete parameter
dependence required for the analysis of the star's evolution, encoded
in a two-parameter function that can be numerically computed and tabulated
for different star models. This offers an explicit framework for the
consistent inclusion of the r-mode saturation into a star evolution
analysis and supersedes previously necessary model assumptions \cite{Owen:1998xg}.
We will analyze the same star models as in our recent companion paper
\cite{Alford:2010fd} and thereby extend this study to the suprathermal
regime. These include neutron stars, hybrid stars and strange stars.
In addition, motivated by the recent observation of a $2M_{\odot}$
star \cite{Demorest:2010bx,Ozel:2010bz} we only study models that
also yield heavy stars. Moreover, similar to the analytic results
in \cite{Alford:2010fd}, we give approximate analytic expressions
for the maximum saturation amplitude that exhibit the detailed parameter
dependence on the equation of state and the transport properties of
dense matter. In addition to the standard fundamental $m\!=\!2$ r-mode
we also study the saturation of higher multipoles and find that they
saturate at significantly lower amplitudes. In this work we concentrate
on the effects of suprathermal bulk viscosity on the damping times
\cite{Alford:2010ep} and the suitably extended concept of the instability
regions, and defer a comprehensive analysis of the star evolution
to future work.

\section{Damping of star oscillations}

In this section we discuss the prerequisites needed for the analysis
of compact star oscillations and their damping. In particular we discuss
the large amplitude enhancement of the bulk viscosity which will provide
a mechanism for the saturation of r-modes. This topic had been studied
in \cite{Alford:2010gw} and we refer the reader to this work for
further details. The analysis further requires the stable equilibrium
configuration of the star and the r-mode profile. These have been
discussed in more detail in our companion article on the damping of
small amplitude r-modes \cite{Alford:2010fd} and we will here only
briefly recall these results.

\subsection{Suprathermal viscosity}

The bulk viscosity describes the local dissipation of energy in a
fluid element in one cycle of compression and rarefaction, driven
by some oscillation mode of the star. The integration over the whole
star yields the corresponding total energy dissipation of an oscillation
mode as will be discussed in section \ref{sec:Damping-time-scales}.
Recently the bulk viscosity of large-amplitude oscillations has been
studied in detail in \cite{Alford:2010gw}, which builds on the classic
work \cite{Madsen:1992sx} and provides general expressions valid
for various forms of matter and arbitrary equations of state.

Bulk viscosity is generally induced by slow weak-interaction processes
whose rate takes the parametric form

\begin{equation}
\Gamma^{(\leftrightarrow)}=-\tilde{\Gamma}T^{\delta}\mu_{\Delta}\Biggl(1+\sum_{j=1}^{N}\chi_{j}\left(\frac{\mu_{\Delta}^{2}}{T^{2}}\right)^{j}\Biggr)\,,\label{eq:gamma-parametrization}\end{equation}
where $T$ is the temperature and $\mu_{\Delta}$ represents the
quantity that is driven out of equilibrium due to the oscillations
and its re-equilibration leads to the bulk viscosity. The latter is
given by the difference of the sums of chemical potentials of the
particles in the initial and final state of the relevant weak process.
The $\chi_{j}$ are coefficients that characterize the non-linear
(large amplitude) contribution to the re-equilibration rate. The series
terminates at a finite order $N$ which is determined by the number
and type of particles in the initial and final states of the dominant
re-equilibration process and is connected to the temperature dependence
via $\delta\!=\!2N$.

As has been shown in \cite{Alford:2010gw} the bulk viscosity can
be written in a general form that expresses its full underlying parameter
dependence in terms of the coefficients $d$ of the driving term and
$f$ of the feedback term

\[
d\equiv\frac{C}{T}\frac{\Delta n}{\bar{n}}\:,\quad f\equiv\frac{B\tilde{\Gamma}T^{\delta}}{\omega}\,,\]
in the differential equation that determines the oscillation $\mu_{\Delta}$.
These expressions depend on the oscillation frequency $\omega$, the
conserved number density fluctuation $\Delta n/\bar{n}$, as well
as the susceptibilities

\begin{equation}
C\equiv\bar{n}\left.\frac{\partial\mu_{\Delta}}{\partial n}\right|_{x}\quad,\quad B\equiv\frac{1}{\bar{n}}\left.\frac{\partial\mu_{\Delta}}{\partial x}\right|_{n}\,,\label{eq:susceptibilities}\end{equation}
with respect to the density $n$ and the fraction $x$ of a particular
particle that is driven out of equilibrium. The general result can
be parameterized in the form \cite{Alford:2010gw}

\begin{equation}
\zeta=\zeta_{max}^{<}\,{\cal I}\left(d,f\right)\,,\label{eq:general-viscosity}\end{equation}
where the global maximum of the viscosity

\[
\zeta_{max}^{<}=\frac{C^{2}}{2\omega B}\]
is taken at

\begin{equation}
d=0\:,\negmedspace f=1\quad\Rightarrow\quad T_{max}=\left(\frac{\omega}{\tilde{\Gamma}B}\right)^{\frac{1}{\delta}}\label{eq:maximum-temperature}\end{equation}
and the non-trivial dimensionless integral

\begin{equation}
{\cal I}(d,f)\equiv\frac{2}{\pi Td}\int_{0}^{2\pi}\mu_{\Delta}(\varphi;d,f)\cos(\varphi)d\varphi\label{eq:viscosity-integral}\end{equation}
has to be tabulated for each considered form of matter with a particular
dominant weak re-equilibration process. Since both $d$ and $f$ depend
on the temperature and the required parameter regions are therefore
non-trivial, see \cite{Alford:2010ht}, this is most conveniently
done in terms of the new variable $\tilde{d}\!\equiv\! d\, f^{1/\delta}$
for the function $\tilde{{\cal I}}\left(\tilde{d},f\right)\!=\!{\cal I}(d,f)$.
This form of the bulk viscosity is crucial for the study of damping
times below, which involve an integration over the whole star, and
therefore requires the complete density and amplitude dependence of
the viscosity in addition to the temperature and frequency dependence
which are used in standard analyses.

In general the bulk viscosity features three distinct characteristic
regions. Almost all previous analyses of r-mode damping have been
limited to the \emph{subthermal} regime $\mu_{\Delta}\!\ll\! T$,
where $\mu_{\Delta}$ is linear in $d\!\sim\!\Delta n$, so the viscosity
is independent of the amplitude and has the analytic form

\begin{equation}
\zeta^{<}=\zeta_{max}^{<}\frac{2f}{1+f^{2}}=\frac{C^{2}\tilde{\Gamma}T^{\delta}}{\omega^{2}+(B\tilde{\Gamma}T^{\delta})^{2}}\label{eq:sub-viscosity}\end{equation}
Yet, since the r-mode is unstable and rises exponentially it eventually
reaches the \emph{suprathermal regime} $\mu_{\Delta}\!\gtrsim\! T$
which will be studied below. This regime is divided into two qualitatively
different parts. In an intermediate regime $\mu_{\Delta}$ is still
linear in $d$ but the viscosity strongly rises. For $f\!\ll\!1$
this part allows an analytic solution given by \begin{align}
\zeta^{\sim} & =\zeta_{max}^{<}f\sum_{j=0}^{N}\frac{\left(2j+1\right)!!\chi_{j}}{2^{j-1}\left(j+1\right)!}d^{2j}\label{eq:madsen-approximation}\\
 & =\frac{C^{2}\tilde{\Gamma}T^{\delta}}{\omega^{2}}\left(1+\sum_{j=1}^{N}\frac{\left(2j+1\right)!!\chi_{j}}{2^{j}\left(j+1\right)!}\biggl(\frac{C}{T}\frac{\Delta n}{\bar{n}}\biggr)^{\!2j}\right)\nonumber \end{align}
which can be combined with the subthermal result to give an approximate
analytic solution for both regions $\zeta^{\lesssim}\!=\!\zeta^{<}\!+\!\theta(T\!-\! T_{max})\zeta^{\sim}$.
At even higher amplitudes the rise of $\mu_{\Delta}$ becomes weaker
due to non-linear saturation effects so that the viscosity has a maximum
and decreases again. In the asymptotic limit $\mu_{\Delta}\!\gg\! T$,
the viscosity scales as\begin{equation}
\zeta\sim\left(\frac{\Delta n}{\bar{n}}\right)^{-\frac{2N}{2N+1}}\,.\label{eq:supra-limit}\end{equation}
As discussed in appendix \ref{sec:strange-matter-approximation},
for the special case of strange quark matter an approximate analytic
result is possible that includes this large amplitude regime.

In contrast to the bulk viscosity, the shear viscosity of dense matter
is independent of the frequency and amplitude of an external oscillation
and over certain temperature ranges its dependence on temperature
is approximately a simple power law. Shear viscosity becomes large
at low temperatures and therefore it is the dominant process for damping
of the r-modes of cooler stars. Correspondingly we can be parameterize
the shear viscosity as

\begin{equation}
\eta=\tilde{\eta}T^{-\sigma}\label{eq:shear-parametrization}\end{equation}
by simply factoring out the temperature dependence with exponent $\sigma$.

\subsection{Star models}

We study in this work the same model examples of compact stars as
in our companion paper \cite{Alford:2010fd}. These include neutron
stars, strange stars \cite{Witten:1984rs}, and hybrid stars, and
we consider in each case a star with a standard mass of $1.4\, M_{\odot}$
and a heavy star with a mass of $2\, M_{\odot}$. For the neutron
stars we use nuclear matter obeying the APR equation of state \cite{Akmal:1998cf}
and as a low density extension of the APR data we use \cite{Baym:1971pw,Negele:1971vb}.
In order to apply our general results to other equations of state,
away from chemical equilibrium we use the simple quadratic parameterization
in terms of the symmetry energy employed in \cite{Lattimer:1991ib}.
With the exception of ultra-heavy neutron stars the APR equation of
state allows only modified Urca processes \cite{Friman:1978zq,Sawyer:1989dp,Reisenegger:1994be}
which are the dominant microscopic processes that induce bulk viscosity.
Direct Urca processes \cite{Haensel:1992zz,Reisenegger:1994be} are
only possible in the core of a neutron star close to the mass limit
which we study in addition. The dominant contribution to the shear
viscosity in hadronic matter results from non-Fermi liquid enhanced
lepton-lepton scattering \cite{Shternin:2008es}. 

For the strange stars and the quark core of the hybrid stars we use
quark matter obeying a simple equation of state in terms of parameters
$c$, $m_{s}$ and $B$,

\begin{align}
p_{par} & =\frac{1-c}{4\pi^{2}}\left(\mu_{d}^{4}+\mu_{u}^{4}+\mu_{s}^{4}\right)-\frac{3m_{s}^{2}\mu_{s}^{2}}{4\pi^{2}}\label{eq:quark-eos-model}\\
 & \quad+\frac{3m_{s}^{4}}{32\pi^{2}}\left(3+4\log\!\left(\frac{2\mu_{s}}{m_{s}}\right)\right)-{\cal B}+\frac{\mu_{e}^{4}}{12\pi^{2}}\,.\nonumber \end{align}
This is a generalization of the parameterization employed in \cite{Alford:2004pf}
from which the equilibrium pressure is obtained via the conditions
of weak equilibrium $\mu_{s}\!=\!\mu_{d}\!=\!\mu_{u}\!+\!\mu_{e}$
and electrical neutrality. The shear viscosity in quark matter arises
from non-Fermi-liquid-enhanced quark-quark scattering \cite{Heiselberg:1993cr}
and the bulk viscosity from non-leptonic flavor-changing weak processes
\cite{Heiselberg:1992bd,Madsen:1992sx}. For our hybrid stars we make
the assumption of local electrical neutrality, excluding the possibility
of a mixed phase and its wealth of geometric structures.

The equilibrium star configuration is then determined by the general-relativistic
Tolman-Oppenheimer-Volkov (TOV) equations \cite{Tolman:1939jz}. Characteristics
of the examples that we consider are given in table \ref{tab:star-models}.
The susceptibilities $B$ and $C$ of the considered forms of matter
are given in table \ref{tab:strong-parameters}. Finally, the reequilibration
parameters $\tilde{\Gamma}$, $\delta$ and $\chi$ in the parameterization
of the weak rate eq.\,(\ref{eq:gamma-parametrization}) and the parameters
of the shear viscosity eq.\,(\ref{eq:shear-parametrization}) are
given for the considered forms of dense matter in table \ref{tab:weak-parameters}.
Note that there are higher non-linearities in the weak rate of hadronic
compared to quark matter, which in accordance with eq.\,(\ref{eq:madsen-approximation})
yields a steeper rise of the viscosity with amplitude so that suprathermal
effects are even more important in hadronic matter than in quark matter.

\begin{table}[H]
\begin{tabular}{|c|c|c|c|c|c|c|}
\hline 
 & $M\left[M_{\odot}\right]$ & $M_{core}\left[M_{\odot}\right]$ & $R\left[km\right]$ & $n_{c}\left[n_{0}\right]$ & $\left\langle n\right\rangle \left[n_{0}\right]$ & $\Omega_{K}\left[kHz\right]$\tabularnewline
\hline 
NS & $1.4$ & $\left(1.39\right)$ & $11.5$ & $3.43$ & $1.58$ & $6.02$\tabularnewline
\hline 
 & $2.0$ & $\left(1.99\right)$ & $11.0$ & $4.91$ & $2.46$ & $7.68$\tabularnewline
\hline 
 & $2.21$ & $0.85$ & $10.0$ & $7.17$ & $3.37$ & $9.31$\tabularnewline
\hline 
SS & $1.4$ & $-$ & $11.3$ & $2.62$ & $1.91$ & $6.17$\tabularnewline
\hline 
 & $2.0$ & $-$ & $11.6$ & $4.95$ & $2.43$ & $7.09$\tabularnewline
\hline 
HS & $1.4$ & $1.06$ & $12.7$ & $2.32$ & $1.17$ & $5.16$\tabularnewline
\hline 
 & $2.0$ & $1.81$ & $12.2$ & $4.89$ & $1.84$ & $6.62$\tabularnewline
\hline
\end{tabular}

\caption{\label{tab:star-models}Properties of the considered models of neutron
stars (NS), strange stars (SS) and hybrid stars (HS). We show the
mass of the star $M$, the mass of the core $M_{core}$, the radius
$R$, the baryon density at the center of the star $n_{c}$ given
in units of nuclear saturation density $n_{0}$, the average density
$\left\langle n\right\rangle $ and the Kepler frequency $\Omega_{K}$.
The neutron stars were obtained by solving the relativistic TOV equations
for catalyzed neutron matter using the APR equation of state \cite{Akmal:1998cf}
with low density extension \cite{Baym:1971pw,Negele:1971vb} and the
strange stars with a quark gas bag model with $c=0$, $m_{s}=150\,\MeV$
and a bag parameter $B=\left(138\,{\rm \MeV}\right)^{4}$. Large mass
hybrid stars are only found when strong interaction corrections are
considered, cf. \cite{Alford:2004pf}, and we find a $2\, M_{\odot}$
star for $c=0.4$, $m_{s}=140\,\MeV$, $B=\left(137\,\MeV\right)^{4}$.}

\end{table}

\begin{table*}
\begin{tabular}{|c|c|c|c|}
\hline 
 & $A$ & $B$ & $C$\tabularnewline
\hline 
hadronic matter & $m_{N}\left(\frac{\partial p}{\partial n}\right)^{-1}$ & $\frac{8S}{n}\!+\negthinspace\frac{\pi^{2}}{\left(4\left(1\!-\!2x\right)S\right)^{2}}$ & $4\!\left(1\!-\!2x\right)\!\left(n\frac{\partial S}{\partial n}\!-\!\frac{S}{3}\right)$\tabularnewline
\hline 
hadronic gas & $\frac{3m_{N}^{2}}{\left(3\pi^{2}n\right)^{2/3}}$ & $\frac{4m_{N}^{2}}{3\left(3\pi^{2}\right)^{1/3}n^{4/3}}$ & $\frac{\left(3\pi^{2}n\right)^{2/3}}{6m_{N}}$\tabularnewline
\hline 
quark matter (gas: $c=0$) & $3\!+\!\frac{m_{s}^{2}}{\left(1\!-\! c\right)\mu_{q}^{2}}$ & $\frac{2\pi^{2}}{3(1-c)\mu_{q}^{2}}\left(1\!+\!\frac{m_{s}^{2}}{12(1-c)\mu_{q}^{2}}\right)$ & $-\frac{m_{s}^{2}}{3(1-c)\mu_{q}}$\tabularnewline
\hline
\end{tabular}

\caption{\label{tab:strong-parameters}Strong interaction parameters, defined
in eqs.\,(\ref{eq:susceptibilities}) and (\ref{eq:A-parameter}),
describing the response of the particular form of matter. In the case
of interacting hadronic matter a quadratic ansatz in the proton fraction
$x$ parameterized by the symmetry energy $S$ is employed. The expressions
for a hadron and quark gas are given to leading order in $n/m_{N}^{3}$
respectively next to leading order in $m_{s}/\mu$.}

\end{table*}
\begin{table*}
\begin{minipage}[t]{0.6\textwidth}%
\begin{tabular}{|c||c|c|c|c|c|}
\hline 
Weak process & $\tilde{\Gamma}\,\bigl[\mathrm{MeV}^{(3-\delta)}\bigr]$ & $\delta$ & $\chi_{1}$ & $\chi_{2}$ & $\chi_{3}$\tabularnewline
\hline
\hline 
quark non-leptonic & $6.59\!\times\!10^{-12}\,\Bigl(\frac{\mu_{q}}{300\,\mathrm{MeV}}\Bigr)^{5}$ & $2$ & $\frac{1}{4\pi^{2}}$ & $0$ & $0$\tabularnewline
\hline 
hadronic direct Urca & $5.24\!\times\!10^{-15}\left(\!\frac{x\, n}{n_{0}}\!\right)^{\!1/3}$ & $4$ & $\frac{10}{17\pi^{2}}$ & $\frac{1}{17\pi^{4}}$ & $0$\tabularnewline
\hline 
hadronic modified Urca & $4.68\!\times\!10^{-19}\left(\!\frac{x\, n}{n_{0}}\!\right)^{\!1/3}$ & $6$ & $\frac{189}{367\pi^{2}}$ & $\frac{21}{367\pi^{4}}$ & $\frac{3}{1835\pi^{6}}$\tabularnewline
\hline
\end{tabular}%
\end{minipage}%
\begin{minipage}[t]{0.4\textwidth}%
\begin{tabular}{|c||c|c|}
\hline 
Strong/EM process & $\tilde{\eta}\,\bigl[\mathrm{MeV}^{(3+\sigma)}\bigr]$ & $\sigma$\tabularnewline
\hline
\hline 
quark scattering & $1.98\!\times\!10^{9}\alpha_{s}^{-\frac{5}{3}}\left(\frac{\mu_{q}}{300\,\mathrm{MeV}}\right)\negthickspace^{\frac{14}{3}}$ & $\frac{5}{3}$\tabularnewline
\hline 
leptonic scattering & $1.40\!\times\!10^{12}\left(\frac{x\, n}{n_{0}}\right)\negthickspace^{\frac{14}{9}}$ & $\frac{5}{3}$\tabularnewline
\hline 
nn-scattering & $5.46\!\times\!10^{9}\left(\frac{n}{m_{N}n_{0}}\right)\negthickspace^{\frac{9}{4}}$ & $2$\tabularnewline
\hline
\end{tabular}%
\end{minipage}

\caption{\label{tab:weak-parameters}\emph{Left panel:} Parameters of the general
parameterization of the weak rate eq.\,(\ref{eq:gamma-parametrization})
for different processes of particular forms of matter which determine
the damping due to bulk viscosity. The coefficients $\chi_{i}$ parameterize
the non-linear dependence on the chemical potential fluctuation $\mu_{\Delta}$
arising in the suprathermal regime of the viscosity which is relevant
for large amplitude r-modes. \emph{Right panel:} Parameters arising
in the parameterization eq.~(\ref{eq:shear-parametrization}) of
the shear viscosity for different strong and electromagnetic interaction
processes. The leptonic and quark scatterings arise from a non-Fermi
liquid enhancement due to unscreened magnetic interactions. }

\end{table*}

\subsection{R-mode profile}

R-modes are normal modes of rotating stars and are obtained from a
linear perturbation analysis around the static star configurations
discussed above. We consider a star rotating with angular frequency
$\Omega$. To simplify this analysis it is performed in a Newtonian
approximation and in a slow rotation expansion in $\Omega^{2}/\left(\pi G_{N}\bar{\rho}\right)$,
where $\bar{\rho}$ is the average density of the unperturbed star.
The analysis of the damping due to bulk viscosity strictly requires
an expansion to next to leading order. The oscillation frequency depends
on the frame. For an r-mode with angular dependence $Y_{m+1}^{m}$
the frequency in a frame rotating with the unperturbed star $\omega_{r}$,
which is relevant for microscopic quantities like the viscosity, and
the corresponding frequency $\omega_{i}$, observed by an observer
in an inertial frame, are given by

\begin{equation}
\omega_{r}\equiv\omega=\kappa(\Omega)\Omega\quad,\quad\omega_{i}=\omega_{r}-m\Omega\label{eq:frequency-connection}\end{equation}
where the function $\kappa$ has an analog expansion and reads to
lowest order $\kappa_{0}\!=\!2/\left(m\!+\!1\right)$. For dissipation
via bulk viscosity the relevant quantity is the density compression
due to the r-mode with dimensionless amplitude $\alpha$. It vanishes
to leading order in the slow rotation expansion and reads at next
to leading order \cite{Lindblom:1999yk,Lindblom:1998wf}

\begin{align}
\left|\frac{\Delta n}{\bar{n}}\right| & \approx\sqrt{\frac{4m}{\left(m+1\right)^{3}\left(2m+3\right)}}\frac{2}{\left(m+1\right)\kappa(\Omega)}\alpha AR^{2}\Omega^{2}\label{eq:r-mode-profile}\\
 & \quad\cdot\left(\left(\left(\frac{r}{R}\right)^{m+1}+\delta\Phi_{0}\right)\left|Y_{m+1}^{m}(\theta,\phi)\right|+\cdots\right)\nonumber \end{align}
where the ellipsis denotes further contributions involving next to
leading order corrections of the potentials that are explicitly given
in \cite{Lindblom:1999yk}. In this expression $A$ denotes the inverse
squared speed of sound

\begin{equation}
A\equiv\left.\frac{\partial\rho}{\partial p}\right|_{0}\label{eq:A-parameter}\end{equation}
evaluated at equilibrium. There are different conventions for the
amplitude $\alpha$ in the literature and we follow the convention
for the amplitude given in \cite{Lindblom:1998wf} but take into account
the corrections to the latter result in \cite{Lindblom:1999yk}. This
convention is usually used in the literature %
\footnote{We note that we had previously used the alternative $\alpha$-definition
\cite{Lindblom:1999yk} in the proceedings article \cite{Alford:2010ep}
and a preprint version of the present article, which led to lower
values for the saturation amplitudes, see also Appendix \ref{sec:R-mode-profile}.%
} and in this case the above expression breaks down for $\alpha\!>\! O\left(1\right)$.
For more details on the r-mode expression see Appendix \ref{sec:R-mode-profile}.
In general one must numerically solve a differential equation to obtain
$\delta\Phi_{0}$, which is the leading order correction to the gravitational
potential \cite{Alford:2010fd}. However, in the special case of a
star with a constant density profile $\delta\Phi_{0}$ can be shown
to be subleading compared to the first term in the inner parentheses
of eq.\,(\ref{eq:r-mode-profile}). For our numerical analysis below
we will make the approximation to neglect the additional second order
corrections in the slow rotation expansion given by the ellipsis in
eq. (\ref{eq:r-mode-profile}), which amounts to replacing the Lagrangian
perturbation by the Eulerian perturbation, but we include the second
order corrections to the frequency eq.\,(\ref{eq:frequency-connection}).
General analytic results showed that this is a good approximation
for the computation of the small amplitude instability regions \cite{Alford:2010fd}.
Here we will give semi-analytic results for the saturation amplitudes,
valid beyond leading order, that show the influence of the second
order terms. 

\begin{figure}
\includegraphics{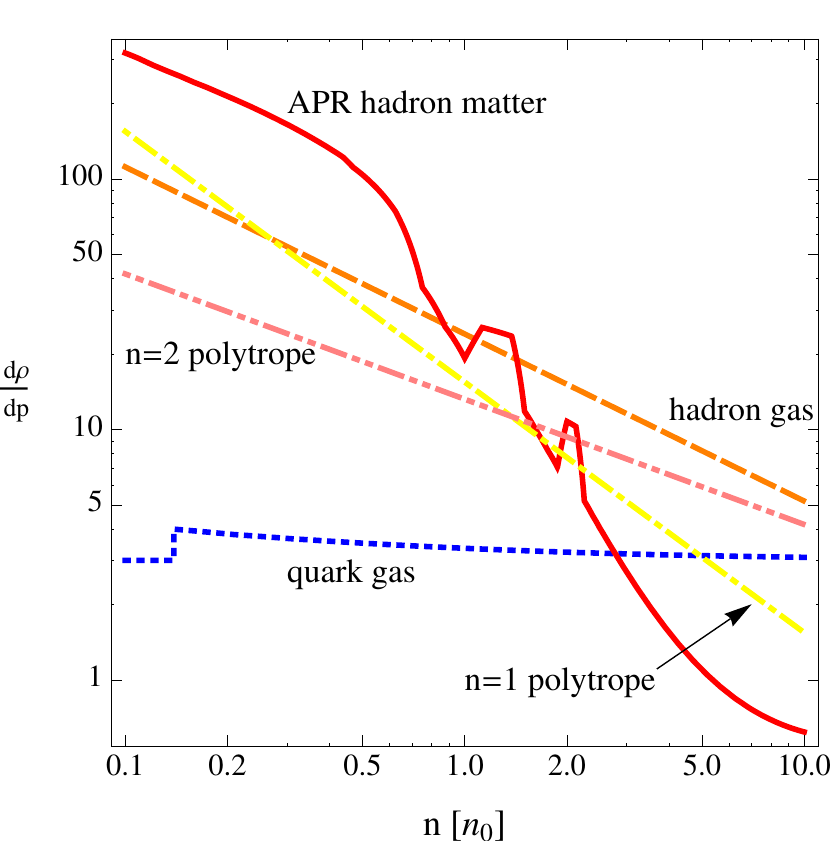}

\caption{\label{fig:dE/dp-density-dependence}The density dependence of the
inverse squared speed of sound $A\equiv d\rho/dp$ (which enters the
r-mode profile multiplicatively) for the different forms of matter
in table\,\ref{tab:strong-parameters} as well as generic polytropic
models. The solid line represents interacting APR matter, the dashed
line a hadron gas and the dotted line shows the result for a quark
gas. The structure at intermediate densities in the APR curve arises
from phase transitions and the use of finite differences to compute
the derivative, but due to the mild contribution of the denser inner
regions of the star to the damping these, as well as the known problem
that the APR equation of state becomes acausal at high density, have
no influence on our results below.}

\end{figure}

We will see below that, in obtaining a precise assessment of the damping
time of large amplitude r-modes, the radial dependence of the density
perturbation plays a vital role. The radius enters eq.\,(\ref{eq:r-mode-profile})
explicitly and also via the density dependence of the inverse squared
speed of sound $A$ (fig.\,\ref{fig:dE/dp-density-dependence}) and
the radial density dependence of the star (fig.\,2 in \cite{Alford:2010fd}).
The radial variation of the density is moderate in a strange star,
but much more pronounced in neutron stars where the r-mode amplitude
grows strongly in the outer regions of the star. 

\section{Damping time scales\label{sec:Damping-time-scales}}

\subsection{General expressions}

The amplitude of the r-mode oscillations evolves with time dependence
$\exp(i\omega t-t/\tau)$. We can decompose the decay rate $1/\tau$
as \begin{equation}
\frac{1}{\tau(\Omega)}=\frac{1}{\tau_{G}(\Omega)}+\frac{1}{\tau_{B}(\Omega)}+\frac{1}{\tau_{S}(\Omega)}\end{equation}
 Where $\tau_{G}$, $\tau_{B}$ and $\tau_{S}$ are gravitational
radiation, bulk viscosity and shear viscosity time scales, respectively.
The time scale of the r-mode growth due to gravitational wave emission
is given by \cite{Lindblom:1998wf} 

\begin{equation}
\frac{1}{\tau_{G}}=-\frac{32\pi\left(m-1\right)^{2m}}{\left(\left(2m+1\right)!!\right)^{2}}\left(\frac{m+2}{m+1}\right)^{\!2m+2}\tilde{J_{m}}GMR^{2m}\Omega^{2m+2}\label{eq:gravitational-time}\end{equation}
and the damping time due to the shear viscosity can be written as
\cite{Alford:2010fd}

\begin{equation}
\frac{1}{\tau_{S}}=\sum_{l}\frac{\left(m-1\right)\left(2m+1\right)\tilde{S}_{m}^{\left(l\right)}\Lambda_{{\rm QCD}}^{3+\sigma}R}{\tilde{J}_{m}MT^{\sigma}}\label{eq:shear-viscosity-damping-time}\end{equation}
with the radial integral constants

\begin{align}
\tilde{J_{m}} & \equiv\frac{1}{MR^{2m}}\int_{0}^{R}\rho\left(r\right)r^{2m+2}dr\label{eq:J-tilde}\\
\tilde{S}_{m}^{\left(l\right)} & \equiv\frac{1}{R^{2m+1}\Lambda_{{\rm QCD}}^{3+\sigma}}\int_{R_{i}^{\left(l\right)}}^{R_{o}^{\left(l\right)}}\tilde{\eta}r^{2m}dr\label{eq:S-tilde}\end{align}
where $\rho$ is the energy density and $\Lambda_{{\rm QCD}}$ a generic
QCD scale introduced to make the constant dimensionless. When several
shells with inner radii $R_{i}^{\left(l\right)}$and outer radii $R_{o}^{\left(l\right)}$
and distinct phases and/or transport coefficients are present in a
compact star, the damping time integral over the star decomposes into
a sum of contributions of the individual shells $l$. The bulk viscosity
damping time is given by \cite{Lindblom:1999yk}

\begin{align}
\frac{1}{\tau_{B}}= & \frac{\kappa^{2}}{\alpha^{2}\tilde{J}_{m}MR^{2}}\int d^{3}x\Bigl|\frac{\Delta\rho}{\bar{\rho}}\Bigr|^{2}\zeta\Bigl(\Bigl|\frac{\Delta\rho}{\bar{\rho}}\Bigr|^{2}\Bigr)\label{eq:bulk-damping-time}\end{align}
Using the general expression for the bulk viscosity eqs.\,(\ref{eq:general-viscosity})
and (\ref{eq:viscosity-integral}) and expressing the fluctuation
in the conserved energy density $\Delta\rho/\bar{\rho}$ by the same
fluctuation of the baryon density $\Delta n/\bar{n}$, gives for the
damping time the general expression

\begin{align}
\frac{1}{\tau_{B}} & =\frac{4\pi\Omega^{3}}{(m+1)^{2}\tilde{J}_{m}MR^{2}\kappa}\sum_{l}{\cal T}^{\left(l\right)}(a,b)\,.\label{eq:general-bulk-viscosity-time}\end{align}
in terms of integrals over the individual shells. Defining the reduced
density oscillation

\begin{equation}
\Bigl(\frac{\Delta n}{\bar{n}}\Bigr)_{\mathrm{red.}}\equiv\frac{\left(m+1\right)\kappa(\Omega)}{2\alpha\Omega^{2}}\frac{\Delta n}{\bar{n}}\end{equation}
we can write 

\begin{align}
 & {\cal T}^{\left(l\right)}(a,b)\equiv\int_{R_{i}^{\left(l\right)}}^{R_{o}^{\left(l\right)}}\! dr\, r^{2}\int\! d\theta\sin\theta\left|\Bigl(\frac{\Delta n}{\bar{n}}\Bigr)_{\mathrm{red.}}(r,\theta)\right|^{2}\frac{C(r)^{2}}{B(r)}\nonumber \\
 & \cdot\tilde{{\cal I}}\left(aC(r)B(r)^{1/\delta}\tilde{\Gamma}(r)^{1/\delta}\left|\Bigl(\frac{\Delta n}{\bar{n}}\Bigr)_{\mathrm{red.}}(r,\theta)\right|,\, bB(r)\tilde{\Gamma}(r)\right)\label{eq:T-function}\end{align}
depending on only two independent parameters

\begin{align*}
a & \equiv\frac{\kappa_{0}\alpha\Omega^{2}}{\kappa\omega^{1/\delta}}=\frac{2\alpha\Omega^{\left(2\delta-1\right)/\delta}}{\left(m+1\right)\kappa^{\left(\delta+1\right)/\delta}}\\
b & \equiv\frac{T^{\delta}}{\omega}=\frac{T^{\delta}}{\kappa\Omega}\end{align*}
Note that in eq.\,(\ref{eq:T-function}) all local quantities can
have different functional forms in different shells, as given in tables
\ref{tab:strong-parameters} and \ref{tab:weak-parameters}, but to
make the expression readable we do not show the explicit suffixes
$(l)$. Whereas strange stars are basically homogeneous and consist
of a single phase, the crust of neutron and hybrid stars is extremely
inhomogeneous and complicated. Although there are no free protons
and thereby no Urca processes, the ultra-heavy nuclei present in the
inner crust as well as the clusters in potential pasta phases still
feature analogous beta-processes. Since oscillations likewise push
the system out of beta-equilibrium an analogous suprathermal enhancement
of the bulk viscosity contribution from these phases is expected.
However, there are to our knowledge no results for the bulk viscosity
in the inner crust, yet \cite{Chamel:2008ca}. Therefore we will in
our numeric results given below neglect the contribution from the
crust and only include the contribution from the core. The core does
not have a sharply-defined boundary: we chose it conventionally to
be at baryon density $n_{0}/4$ corresponding to the lowest point
in the APR table, but check the dependence on this choice.

\subsection{Approximate limits of the bulk viscosity damping time}

In the subthermal regime $\mu_{\Delta}\!\ll\! T$ the bulk viscosity
eq.\,(\ref{eq:sub-viscosity}) is independent of the r-mode amplitude,
so the angular integral in eq.\,(\ref{eq:bulk-damping-time}) is
trivial. The damping time in the subthermal regime is then given by

\begin{equation}
\frac{1}{\tau_{B}^{<}}=\frac{16m}{(2m+3)(m+1)^{5}\kappa}\frac{R^{5}\Omega^{3}}{\tilde{J}_{m}M}\sum_{l}{\cal T}_{m}^{<\left(l\right)}\Bigl(\frac{T^{\delta}}{\kappa\Omega}\Bigr)\label{eq:subthermal-damping-time}\end{equation}
in terms of the one dimensional radial integral\[
{\cal T}_{m}^{<\left(l\right)}(b)\equiv\frac{b}{R^{3}}\int_{R_{i}^{\left(l\right)}}^{R_{o}^{\left(l\right)}}dr\, r^{2}\frac{A^{2}C^{2}\tilde{\Gamma}}{1+\tilde{\Gamma}^{2}B^{2}b^{2}}\left(\left(\frac{r}{R}\right)^{m+1}+\delta\Phi_{0}\right)^{2}\]
This expression was used to study the small amplitude instability
regions in \cite{Alford:2010fd}. Here we want to study the large-amplitude
saturation and therefore it is useful to obtain an analytic expression
that includes the large-amplitude enhancement of the bulk viscosity.
In the intermediate, linear regime and for $f\!\ll\!1$ the general
analytic approximation for the bulk viscosity eq.\,(\ref{eq:madsen-approximation})
is valid. Since this local condition has to be fulfilled everywhere
in the star, the global parameters $a$ and $b$ must be smaller than
certain bounds that are determined by the particular properties of
the considered stars. We recall from \cite{Alford:2010gw} that the
approximation is particularly useful for hadronic matter with modified
Urca processes where it covers almost the entire range of physical
local density amplitudes at millisecond frequencies. A plot of the
regions of validity of the individual analytic approximations of the
bulk viscosity for different forms of matter is given in \cite{Alford:2010ht}.
Analogous to the low temperature/high frequency approximation in the
subthermal regime eq. (\ref{eq:subthermal-damping-time}), in the
intermediate, linear regime an explicit evaluation is possible. With
the analytic expression for the bulk viscosity eq. (\ref{eq:madsen-approximation})
the angular integrals over the spherical harmonics arising in the
r-mode profile take the form

\[
\int\!\! d\Omega_{\theta\phi}\left|Y_{m+1}^{m}(\theta,\phi)\right|^{2n}\!=\!\frac{4\pi\left(2n-1\right)!!(m\, n)!}{(2(m+1)n+1)!!}\Bigl(\frac{\left(2m+3\right)!!}{4\pi\, m!}\Bigr)^{n}\]
and this yields a result that apart from the evaluation of the remaining
radial integrals is analytic (see also \cite{Reisenegger:2003pd,Bonacic:2003th})

\begin{widetext}

\begin{equation}
\frac{1}{\tau_{B}^{\sim}}=\frac{16\left(2m+1\right)!!\Lambda_{QCD}^{9-\delta}R^{5}T^{\delta}\Omega^{2}}{(m+1)^{5}(m-1)!\kappa^{2}\tilde{J}_{m}\Lambda_{EW}^{4}M}\sum_{j=0}^{N}\frac{((2j+1)!!)^{2}(m(j+1))!\chi_{j}\tilde{V}_{m,j}}{(j+1)!\left(2(m+1)(j+1)+1\right)!!}\left(\frac{2m\left(2m+1\right)!!}{\pi\left(m+1\right)^{5}m!\kappa^{2}}\frac{\Lambda_{QCD}^{2}R^{4}\alpha^{2}\Omega^{4}}{T^{2}}\right)^{j}\label{eq:linear-bulk-vicosity-time}\end{equation}
\end{widetext}The dependence on all local parameters, like the equation
of state, the weak rate, the density dependence of the particular
star  and its r-mode profile is contained in a few dimensionless radial
integral constants ($\Lambda_{EW}$ is a typical electroweak scale)

\begin{align}
 & \tilde{V}_{m,j}\equiv\frac{\Lambda_{EW}^{4}}{\Lambda_{QCD}^{7-\delta+2(j+1)}R^{3}}\label{eq:V-tilde}\\
 & \cdot\int_{0}^{{\cal R}}\!\! dr\, r^{2}\tilde{\Gamma}\!\left(r\right)\!\biggl(A\!\left(r\right)C\!\left(r\right)\biggl(\left(\frac{r}{R}\right)^{m+1}\!+\!\delta\Phi_{0}\!\left(r\right)\biggr)\biggr)^{2(j+1)}\nonumber \end{align}
but the dependence on the parameters of the r-mode evolution $\Omega$,
$\alpha$ and $T$ is entirely explicit in eq.\,(\ref{eq:linear-bulk-vicosity-time}).
The $j\!=\!0$ term in eq.\,(\ref{eq:linear-bulk-vicosity-time})
is precisely the approximate subthermal result eq.\,(\ref{eq:subthermal-damping-time})
in the considered approximation. The constants $\tilde{V}_{j}\!\equiv\!\tilde{V}_{2,j}$
for the fundamental r-mode are given for several stars in table \ref{tab:parameters-amplitude-values}.
Although these parameters can vary significantly for different stars,
it is quite striking that as far as the bulk viscosity is concerned,
the complex details of the individual stars are encoded in a few constants.
Note in particular that the parametric form eq.\,(\ref{eq:linear-bulk-vicosity-time})
remains valid for the full second order r-mode expression and only
the constants eq.\,(\ref{eq:V-tilde}) are changed. At sufficiently
large amplitudes the largest power in the sum in eq.\,(\ref{eq:linear-bulk-vicosity-time})
dominates and due to the connection $\delta\!=\!2N$ the bulk viscosity
damping time becomes temperature independent in this approximation.
Note also that the integrals $\tilde{V}_{m,N}$, as well as the general
expression eq.\,(\ref{eq:T-function}) feature an extremely pronounced
radial dependence, both due to the explicit radial dependence and
the radial dependence of the inverse squared speed of sound $A$,
that strongly weights the outer parts of the star.

\begin{table*}
\begin{tabular}{|c|c|c|c|c|c|c|c|c|}
\hline 
star model & shell & $\tilde{J}$ & $\tilde{S}$ & $\tilde{V}_{0}$ & $\tilde{V}_{1}$ & $\tilde{V}_{2}$ & $\tilde{V}_{3}$ & $\alpha_{sat}\left(\Omega_{K}\right)$\tabularnewline
\hline 
NS $1.4\, M_{\odot}$ & core & $1.81\times10^{-2}$ & $7.68\times10^{-5}$ & $1.31\times10^{-3}$ & $4.24\times10^{-3}$ & $2.02\times10^{-2}$ & $0.105$ & $3.68$\tabularnewline
\cline{1-1} \cline{4-9} 
NS $1.4\, M_{\odot}$ gas &  &  & $4.32\times10^{-6}$ & $1.28\times10^{-4}$ & $5.52\times10^{-5}$ & $3.88\times10^{-5}$ & $3.03\times10^{-5}$ & $14.3$\tabularnewline
\cline{1-1} \cline{3-9} 
NS $2.0\, M_{\odot}$ &  & $2.05\times10^{-2}$ & $2.25\times10^{-4}$ & $1.16\times10^{-3}$ & $4.92\times10^{-3}$ & $3.25\times10^{-2}$ & $0.238$ & $2.52$\tabularnewline
\hline 
NS $2.21\, M_{\odot}$ & d.U. core & $2.02\times10^{-2}$ & $5.05\times10^{-4}$ & $1.16\times10^{-8}$ & $7.24\times10^{-12}$ & $5.39\times10^{-15}$ & $-$ & $-$\tabularnewline
\cline{2-2} \cline{5-9} 
 & m.U. core &  &  & $9.34\times10^{-4}$ & $4.42\times10^{-3}$ & $3.39\times10^{-2}$ & $0.288$ & $2.60$\tabularnewline
\hline
SS eq.\,(\ref{eq:quark-eos-model}) & all & $\frac{3}{28\pi}$ & $\frac{\hat{\eta}\mu_{q}^{14/3}}{5\Lambda_{QCD}^{14/3}\alpha_{s}^{5/3}}$ & $\frac{\Lambda_{EW}^{4}\hat{\Gamma}m_{s}^{4}\mu_{q}^{3}}{9\Lambda_{QCD}^{7}(1-c)^{2}}$ & $\frac{\Lambda_{EW}^{4}\hat{\Gamma}m_{s}^{8}\mu_{q}}{15\Lambda_{QCD}^{9}(1-c)^{4}}$ & $-$ & $-$ & eq.\,(\ref{eq:analytic-quark-sat-amp})\tabularnewline
\cline{1-1} \cline{3-9} 
SS $1.4\, M_{\odot}$ &  & $3.08\times10^{-2}$ & $3.49\times10^{-6}$ & $3.53\times10^{-10}$ & $1.24\times10^{-12}$ & $-$ & $-$ & $1.16$\tabularnewline
\cline{1-1} \cline{3-9} 
SS $2.0\, M_{\odot}$ &  & $2.65\times10^{-2}$ & $4.45\times10^{-6}$ & $3.58\times10^{-10}$ & $9.70\times10^{-13}$ & $-$ & $-$ & $1.56$\tabularnewline
\hline 
HS $1.4\, M_{\odot}$ & quark core & $1.70\times10^{-2}$ & $3.11\times10^{-6}$ & $1.38\times10^{-10}$ & $1.75\times10^{-13}$ & $-$ & $-$ & $2.25$\tabularnewline
\cline{2-2} \cline{4-9} 
 & hadr. core &  & $9.71\times10^{-7}$ & $1.39\times10^{-3}$ & $4.70\times10^{-3}$ & $2.23\times10^{-2}$ & $0.116$ & $3.66$\tabularnewline
\hline 
HS $2.0\, M_{\odot}$ & quark core & $2.00\times10^{-2}$ & $5.25\times10^{-6}$ & $3.76\times10^{-10}$ & $7.75\times10^{-13}$ & $-$ & $-$ & $1.59$\tabularnewline
\cline{2-2} \cline{4-9} 
 & hadr. core &  & $5.24\times10^{-6}$ & $1.07\times10^{-3}$ & $4.12\times10^{-3}$ & $2.31\times10^{-2}$ & $0.134$ & $2.94$\tabularnewline
\hline
\end{tabular}

\caption{\label{tab:parameters-amplitude-values}Radial integral parameters
and static saturation amplitude of a $m\!=\!2$ r-mode for the stars
considered in this work. The constant $\tilde{J}$, $\tilde{S}$ and
$\tilde{V_{i}}$ are given by eqs.\,(\ref{eq:J-tilde}), (\ref{eq:S-tilde})
and (\ref{eq:V-tilde}), respectively, using the generic normalization
scales $\Lambda_{QCD}=1$ GeV and $\Lambda_{EW}=100$ GeV. Note that
the subthermal parameter $\tilde{V}_{0}$ corresponds to $\tilde{V}$
in \cite{Alford:2010fd} where the subscript was omitted for simplicity
and the strange star expressions are given to leading order in $m_{s}/\mu$.}

\end{table*}

\subsection{Results for the damping times}

\begin{figure*}
\begin{minipage}[t]{0.5\textwidth}%
\includegraphics[scale=0.85]{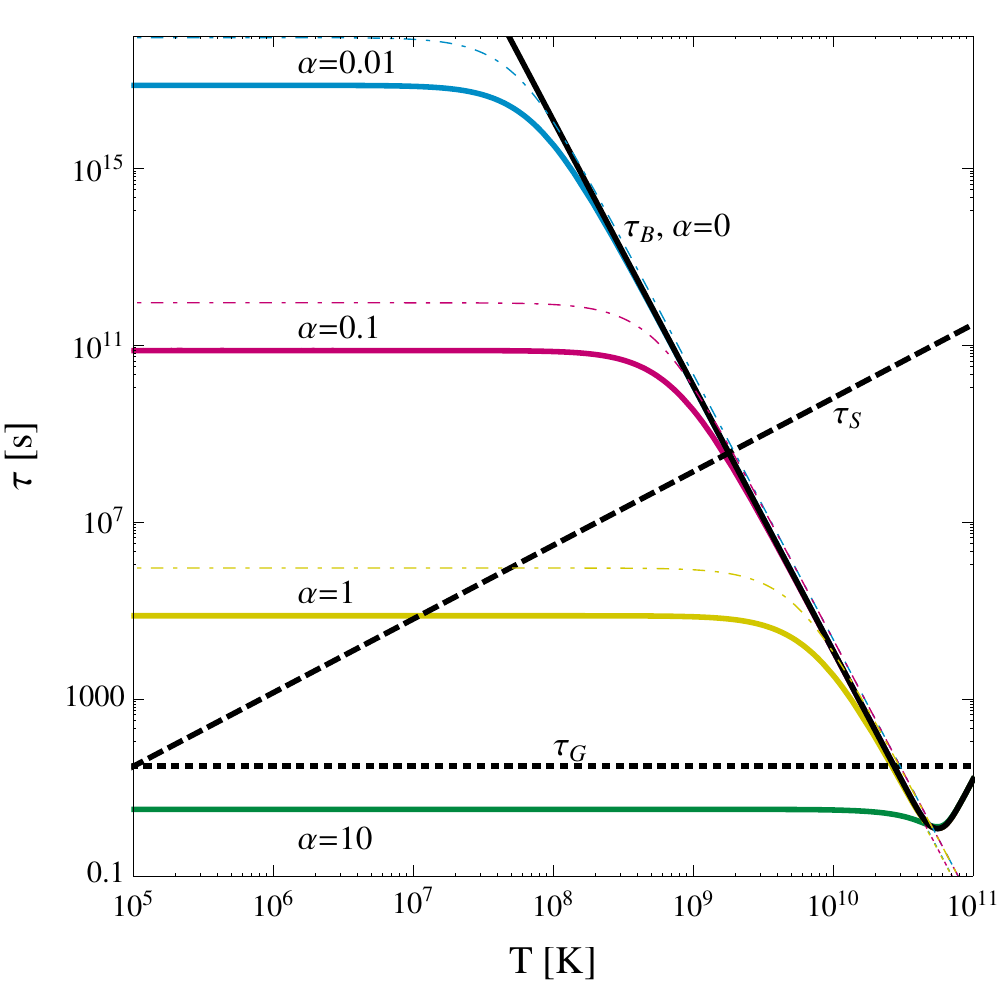}%
\end{minipage}%
\begin{minipage}[t]{0.5\textwidth}%
\includegraphics[scale=0.85]{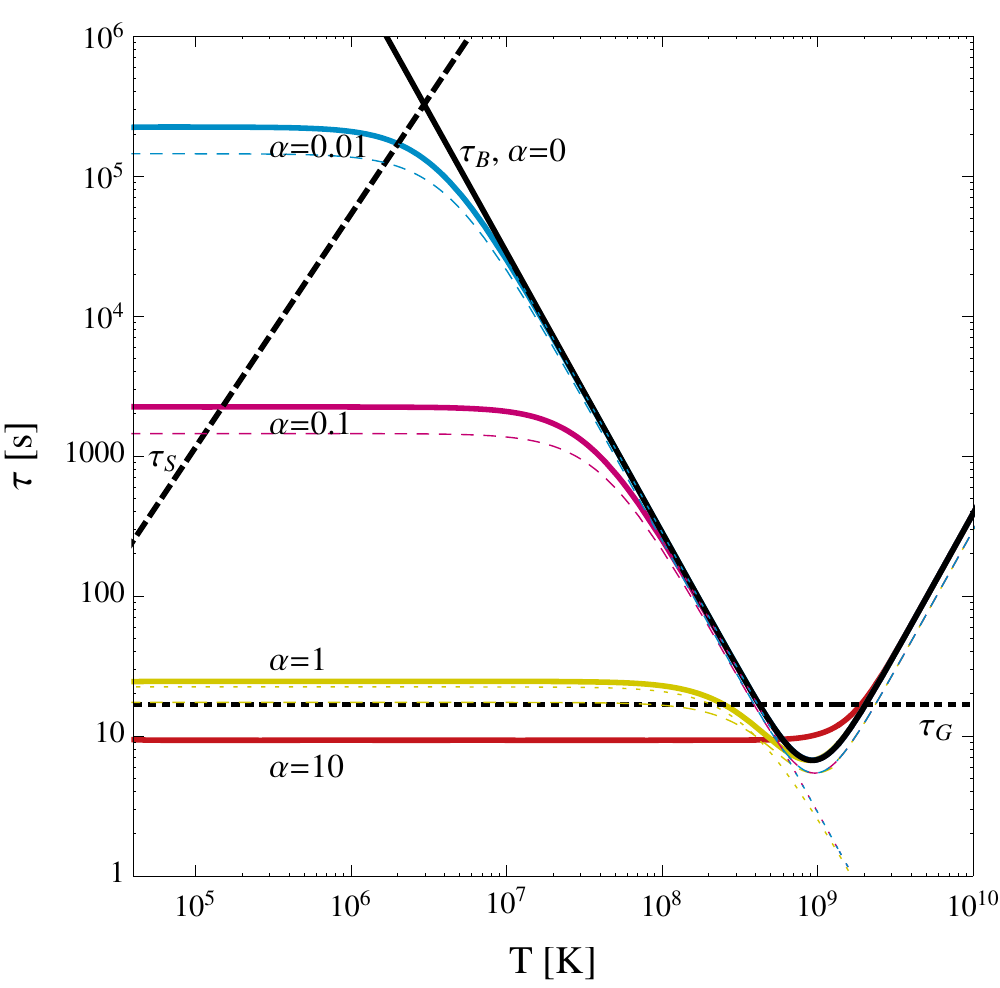}%
\end{minipage}

\caption{\textcolor{green}{\label{fig:damping-times}}The relevant r-mode time
scales for $1.4\, M_{\odot}$ stars rotating at their Kepler frequency.
\emph{Left panel: }Neutron star. \emph{Right panel: }Strange star.
The dotted horizontal line presents the time scale $\tau_{G}$ associated
to the growth of the mode due to gravitational wave emission. The
dashed rising curve shows the damping time $\tau_{S}$ due to shear
viscosity. The damping time $\tau_{B}$ due to bulk viscosity is given
for different dimensionless r-mode amplitudes $\alpha=0$, $0.01$,
$0.1$, $1$ and $10$ by the solid curves. The thin dotted curves
correspond to the analytic linear approximation eq. (\ref{eq:linear-bulk-vicosity-time})
and are below the shown plot range for the largest amplitude. The
thin dot-dashed curves on the left panel show the change when only
a smaller core (ranging to a density of $n_{0}/2$ instead of $n_{0}/4$)
is taken into account. The thin dashed curves on the right panel represent
the approximate analytic expression eq. (\ref{eq:homogenous-quark-damping-time})
given in the appendix which is not valid above the maximum of the
bulk viscosity and therefore not shown for the large amplitude results.}

\end{figure*}

Using the expressions for the microscopic parameters given in tables
\ref{tab:strong-parameters} and \ref{tab:weak-parameters} in the
general expressions eqs. (\ref{eq:gravitational-time}), (\ref{eq:shear-viscosity-damping-time})
and (\ref{eq:general-bulk-viscosity-time}) we obtain the gravitational
and viscosity time scales as a function of their dependent macroscopic
parameters. These are given in fig.\,\ref{fig:damping-times} for
the cases of a neutron star with damping due to modified Urca reactions
(left panel) and a strange star (right panel) by the solid lines as
a function of temperature and for different amplitudes ranging from
top to bottom from the subthermal result at infinitesimal amplitude
to the extreme case $\alpha\!=\!10.$

At sufficiently low temperature the strong increase of the bulk viscosity
with the (local) amplitude $\Delta n/\bar{n}$ damps r-modes with
large (global) dimensionless amplitude $\alpha$ at significantly
shorter time scales. As found before from the analytic expression
eq.\,(\ref{eq:linear-bulk-vicosity-time}), given by the dotted curves,
the damping time is temperature independent in this low temperature
and intermediate amplitude regime. In contrast, due to the generic
form of the bulk viscosity, featuring a universal maximum, the damping
of large amplitude r-modes is not enhanced at high temperatures. As
a direct consequence of the subthermal maximum of the bulk viscosity
\cite{Alford:2010gw}, the corresponding {}``resonant'' temperature
where the damping time is minimal is at roughly $10^{9}$ K for strange
stars and $10^{11}$ K for neutron stars. Correspondingly, r-modes
are entirely unstable at high temperatures. However, due to the strong
suprathermal enhancement at low temperatures the damping undercuts
the gravitational time scale at sufficiently large amplitude so that
the r-mode growth will slow down and eventually saturate. The corresponding
amplitudes are strikingly very similar for the two different classes
of stars, as will be discussed in more detail below. At very large
amplitudes $\alpha\!\sim\! O\left(10\right)$ the damping times decrease
again at all temperatures as a consequence of the behavior of the
bulk viscosity \cite{Alford:2010gw}.

The dot-dashed curves in the neutron star plot on the left panel of
fig.\,\ref{fig:damping-times} show the damping time if the crust
is assumed to start already at the higher density $n_{0}/2$ instead
of $n_{0}/4$ so that only the contribution from the correspondingly
smaller core is taken into account%
\footnote{The range $n_{0}/4$ to $n_{0}/2$ should provide an estimate for
the uncertainty of this boundary. For instance in \cite{Chamel:2008ca}
an intermediate value of $n_{0}/3$ is given.%
}. Although the damping times are larger, as expected, the amplitude
at which the viscous damping can saturate the mode is not drastically
changed, so that our results given below remain qualitatively unchanged
in this case. Actually, when the damping from the crust would be properly
taken into account this should rather enhance the damping and decrease
the r-mode amplitude compared to those obtained in this work. In the
case of the strange star on the right panel of fig.\,\ref{fig:damping-times}
the dashed curves also show the analytic approximation discussed in
appendix \ref{sec:strange-matter-approximation}, where the star is
assumed to be homogeneous. As can be seen, the corresponding expression
eq. (\ref{eq:homogenous-quark-damping-time}) gives an approximate
estimate for the damping time at all temperatures and for amplitudes
up to the maximum of the viscosity, and only fails at higher amplitudes,
where the bulk viscosity cannot saturate the r-mode anymore and where
it is thereby not physically relevant. The deviations compared to
the numeric result stem mainly from the fact that the density in the
strange star is not entirely constant (fig.\,2 in \cite{Alford:2010fd}). 

Previous neutron star analyses \cite{Reisenegger:2003pd,Bonacic:2003th}
have employed an r-mode profile that does not feature the stronger
additional radial dependence due to the low density enhancement of
the inverse squared speed of sound $A$, shown in fig.\,\ref{fig:dE/dp-density-dependence},
for a realistic equation of state. E.g. the r-mode profile given in
eq.\,(6.6) of \cite{Lindblom:2001hd} features roughly a generic
$r^{3}$-dependence. The strong r-dependence in our present treatment,
however, strongly amplifies the damping in the outer regions of the
star. Therefore, the contribution of the crust to the viscous damping
should be relevant and would further decrease the saturation amplitude.
The current restriction of our analysis to the core presents therefore
an upper bound for the saturation amplitude obtained when the damping
of the whole star is considered. The second order effects in contrast
increase the small amplitude instability region \cite{Lindblom:1999yk}
and can correspondingly be expected to likewise increase the saturation
amplitude. A more thorough treatment of all these effects in the future
is clearly desirable.

\section{Saturation amplitudes}

Because of the strong decrease of the viscous damping time due to
the suprathermal enhancement the damping can dominate at sufficiently
large amplitudes. In this case the definition of the instability regions
have to be extended. The latter are standardly defined in the subthermal
regime and are independent of the amplitude. One could extend this
concept by the definition of amplitude dependent instability regions
which would shrink with increasing amplitude. However, since the amplitude
can neither be inferred from observation nor is it a parameter that
can be dialed, but is rather determined dynamically, we refrain from
this possibility and rather introduce the concept of a \emph{static
saturation amplitude}. The latter is defined by the amplitude at which
the r-mode would saturate at fixed temperature and frequency and is
given by the solution of the equation

\begin{equation}
\frac{1}{\tau_{G}(\Omega)}+\sum_{l}\left(\frac{1}{\tau_{S}^{\left(l\right)}(T)}+\frac{1}{\tau_{B}^{\left(l\right)}(T,\Omega,\alpha_{sat})}\right)=0\label{eq:saturation-condition}\end{equation}
where $l$ runs over the contributions from the different shells of
the star. The boundaries of the above mentioned amplitude dependent
instability regions are by definition simply the contour lines $\alpha_{sat}(T,\Omega)\!=$const.
and the boundary of the classic instability region, in particular,
corresponds to $\alpha_{sat}=0$. In case several solutions of eq.\,(\ref{eq:saturation-condition})
exist, only the smallest one is physical and if no solution exists
then viscous damping alone cannot saturate the r-mode according to
this definition. Actually, at the same time the r-mode grows, the
star generally also cools or reheats and spins down respectively up
so that the above amplitudes do not have to be reached. In particular
the star could leave the parameter regions where a saturation according
to the above criterion is not possible before the r-mode can actually
explode.

\subsection{Analytic approximation}

Similar to the analytic expression for the boundary of the instability
region \cite{Alford:2010fd} an analytic expression for the static
saturation amplitude can be obtained. The analytic linear approximation
$\tau_{B}^{\sim}$ applies as long as the r-mode amplitude is small
enough so that the bulk viscosity is sufficiently below its maximum.
When the r-mode amplitude is at the same time large enough that the
highest power in eq. (\ref{eq:linear-bulk-vicosity-time}) dominates,
the damping time becomes temperature independent. Both conditions
are met sufficiently far inside the instability region so that the
damping time simplifies to

\begin{widetext}

\begin{align*}
\frac{1}{\tau_{B}^{\sim}} & \xrightarrow[j=N=\delta/2]{}\frac{2^{4+N}m^{N}\left(\left(2m+1\right)!!\right)^{N+1}\left(\left(2N+1\right)!!\right)^{2}(m\left(N+1\right))!\chi_{N}\tilde{V}_{m,N}}{\pi^{N}\left(m+1\right)^{5\left(N+1\right)}\left(m!\right)^{N}\left(m-1\right)!\left(N+1\right)!\left(2\left(m+1\right)\left(N+1\right)+1\right)!!\kappa^{2N+2}\tilde{J}_{m}}\frac{\Lambda_{{\rm QCD}}^{9}R^{5+4N}\alpha^{2N}\Omega^{4N+2}}{\Lambda_{EW}^{4}M}\end{align*}
At saturation this has to match the gravitational time scale $1/\tau_{B}+1/\tau_{G}=0$
which yields the general result

\begin{align}
\alpha_{sat}= & \left(\frac{\pi^{1+\delta/2}\left(m-1\right)^{2m}\left(m+1\right)^{3+5\delta/2-2m}\left(m+2\right)^{2+2m}\left(\left(m-1\right)!\right)^{1+\delta/2}\left(\frac{\delta}{2}+1\right)!\left(2\left(m+1\right)\left(\frac{\delta}{2}+1\right)+1\right)!!\kappa^{\delta+2}}{2^{\delta/2-1}\left(\left(2m+1\right)!!\right)^{3+\frac{\delta}{2}}\left(\left(\delta+1\right)!!\right)^{2}(m\left(\frac{\delta}{2}+1\right))!\chi_{\frac{\delta}{2}}}\right)^{1/\delta}\nonumber \\
 & \;\cdot\frac{\tilde{J}_{m}^{2/\delta}\Lambda_{EW}^{4/\delta}G^{1/\delta}M^{2/\delta}}{\tilde{V}_{m,\frac{\delta}{2}}^{1/\delta}\Lambda_{QCD}^{9/\delta}R^{2+\left(5-2m\right)/\delta}\Omega^{2-2m/\delta}}\label{eq:analytic-saturation-amplitude}\end{align}

\end{widetext}where $\kappa$ is defined by eq. (\ref{eq:frequency-connection}).
In the cases of strange stars with non-leptonic processes $\delta=2$
and hadronic matter with modified Urca processes $\delta=6$ this
gives for the $m=2$ r-mode

\begin{align}
\alpha_{sat}^{\left(SS\right)} & \approx5.56\cdot10^{-5}\frac{\tilde{J}}{\tilde{V}_{1}^{1/2}}\frac{M_{1.4}}{R_{10}^{5/2}}\approx1.61\frac{\left(1-c\right)^{2}M_{1.4}}{m_{150}^{4}\mu_{300}^{1/2}R_{10}^{5/2}}\label{eq:analytic-quark-sat-amp}\\
\alpha_{sat}^{\left(NS\right)} & \approx10.8\frac{\tilde{J}^{1/3}}{\tilde{V}_{3}^{1/6}}\frac{M_{1.4}^{1/3}}{R_{10}^{13/6}\Omega_{ms}^{4/3}}\nonumber \end{align}
where $\tilde{J}\!\equiv\!\tilde{J}_{2}$ and $\tilde{V}_{i}\!\equiv\!\tilde{V}_{2,i}$
are given for the normalization scales used in table \ref{tab:parameters-amplitude-values}.
Here $m_{150}$, $\mu_{300}$, $M_{1.4}$, $R_{10}$ and $\Omega_{ms}$
are the effective strange quark mass in units of 150 MeV, the quark
chemical potential in units of 300 MeV, the stars mass in units of
$1.4\, M_{\odot}$, the radius in units of $10$ km and the angular
velocity in units of $2\pi$ kHz corresponding to a millisecond pulsar,
respectively. For strange stars the above expression for the intermediate
amplitude bulk viscosity damping time has the same frequency dependence
as the gravitational time scale eq. (\ref{eq:gravitational-time}),
so $\alpha_{sat}$ is basically constant throughout the instability
region. However, it rises with decreasing frequency for neutron stars
where the frequency dependence of the bulk viscosity is weaker. In
contrast to the analytic expressions for the extrema of the instability
region given in \cite{Alford:2010fd}, which are very insensitive
to the microscopic transport parameters, the saturation amplitude
is more sensitive to the suprathermal bulk viscosity parameter $\tilde{V}$.
Whereas the saturation amplitude of neutron stars still depends on
$\tilde{V}$ rather mildly due to the power $1/6$, the saturation
amplitude for strange stars obtained from the generic equation of
state eq.\,(\ref{eq:quark-eos-model}) decreases with the {}``interaction
parameter'' $c$ and even more strongly with the effective strange
quark mass $m_{s}$.

\subsection{Numeric solution}

\begin{figure*}
\begin{minipage}[t]{0.5\textwidth}%
\includegraphics[scale=0.72]{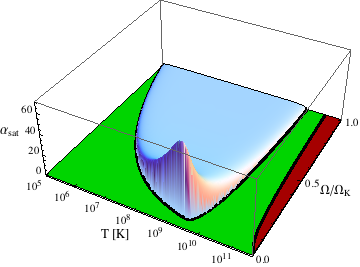}%
\end{minipage}%
\begin{minipage}[t]{0.5\textwidth}%
\includegraphics[scale=0.72]{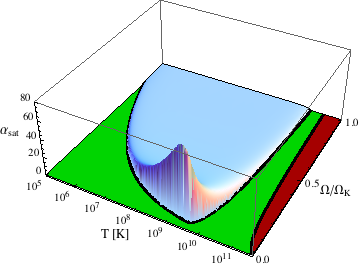}%
\end{minipage}

\caption{\label{fig:saturation-amplitude-APR}The static saturation amplitude,
at which the r-mode growth is stopped by suprathermal viscous damping
for the APR neutron stars. \emph{Left panel:} $1.4\, M_{\odot}$ .
\emph{Right panel:} $2.0\, M_{\odot}$. The light (green) shaded area
denotes the stable region where the r-mode is damped away. At large
frequencies a plateau with amplitudes $O\!\left(1\right)$ is reached.
In the dark (red) region at high temperatures the r-mode is entirely
unstable and cannot be saturated by viscous effects.}

\end{figure*}

\begin{figure*}
\begin{minipage}[t]{0.5\textwidth}%
\includegraphics[scale=0.72]{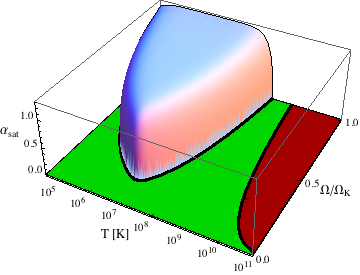}%
\end{minipage}%
\begin{minipage}[t]{0.5\textwidth}%
\includegraphics[scale=0.72]{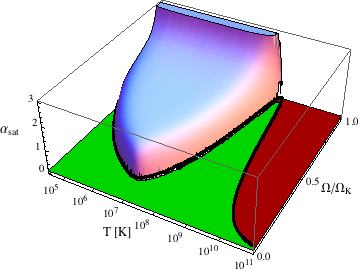}%
\end{minipage}

\caption{\label{fig:saturation-amplitude-Bag}The saturation amplitude for
the considered strange stars. \emph{Left panel:} $1.4\, M_{\odot}$
. \emph{Right panel:} $2.0\, M_{\odot}$ . In the latter case the
suprathermal viscosity cannot stop the r-mode instability at frequencies
larger than the maximum frequency of the stability window (where the
saturation amplitude diverges) - for the considered star slightly
below the Kepler frequency - as well as in the high temperature part
of the instability region. The saturation amplitudes of the plateau
in the lower part of the instability region are of the same order
as in the hadronic case shown in fig.\,\ref{fig:saturation-amplitude-APR}.}

\end{figure*}

\begin{figure*}
\begin{minipage}[t]{0.5\textwidth}%
\includegraphics[scale=0.72]{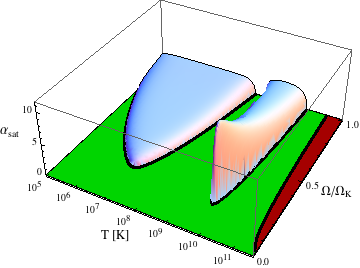}%
\end{minipage}%
\begin{minipage}[t]{0.5\textwidth}%
\includegraphics[scale=0.72]{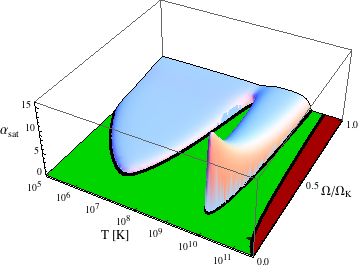}%
\end{minipage}

\caption{\label{fig:saturation-amplitude-Hybrid}The saturation amplitude for
the considered hybrid stars. \emph{Left panel:} $1.4\, M_{\odot}$
. \emph{Right panel:} $2.0\, M_{\odot}$ . The saturation in the low
temperature part of the instability region is mostly established by
the bulk viscosity of the quark core, whereas the saturation in the
mid temperature part comes mainly from the hadronic shell.}

\end{figure*}

\begin{figure*}
\begin{minipage}[t]{0.5\textwidth}%
\includegraphics[scale=0.72]{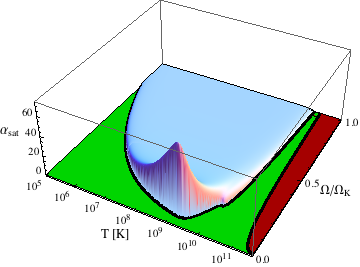}%
\end{minipage}%
\begin{minipage}[t]{0.5\textwidth}%
\includegraphics[scale=0.72]{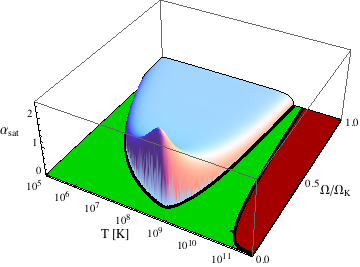}%
\end{minipage}

\caption{\label{fig:saturation-amplitude-APR-dir}The static saturation amplitude,
at which the r-mode growth is stopped by suprathermal viscous damping
for the APR neutron stars at the maximum mass $2.21\, M_{\odot}$,
where direct Urca processes become allowed. \emph{Left panel:} direct
Urca is only allowed in a small inner core region, see Table \ref{tab:star-models}.
\emph{Right panel:} the same model when direct Urca is artificially
turned on in the entire core.}

\end{figure*}

\begin{figure*}
\begin{tabular}{cc}
\includegraphics[scale=0.65]{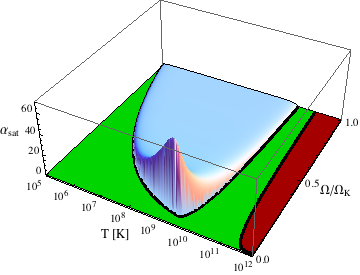} & \includegraphics[scale=0.625]{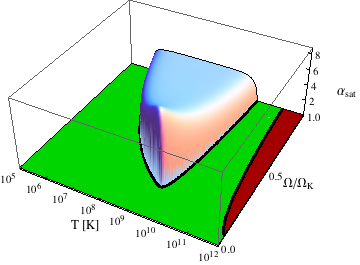}\tabularnewline
\includegraphics[scale=0.64]{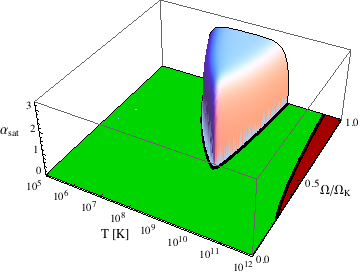} & \includegraphics[scale=0.65]{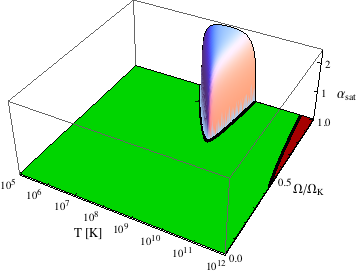}\tabularnewline
\end{tabular}

\caption{\label{fig:instability-regions-multipoles}Saturation amplitudes for
the first four multipole r-modes of the $1.4\, M_{\odot}$ neutron
star  (top, left: $m=2$; top, right: $m=3$; bottom, left: $m=4$;
bottom right: $m=5$). The results are obtained in the linear approximation
eq. (\ref{eq:linear-bulk-vicosity-time}).}

\end{figure*}

Let us now discuss the numerical solution for the saturation amplitude.
In the following plots figs.\,\ref{fig:saturation-amplitude-APR}
to \ref{fig:instability-regions-multipoles} the static saturation
amplitude is shown as a function of temperature and amplitude and
they feature generally 3 distinct regions. The light (blue) surface
shows the saturation amplitude where the r-mode growth is stopped
by suprathermal damping. Due to the characteristic behavior of the
bulk viscosity \cite{Alford:2010gw} which does not feature an amplitude
enhancement for temperatures above the temperature $T_{max}$ eq.
(\ref{eq:maximum-temperature}), the r-mode is not damped at all by
viscous effects in the high temperature regime as denoted by the dark
(red) area on the right hand side. In the flat (green) region surrounding
the instability region the r-mode is entirely stable and already damped
by the shear or the subthermal bulk viscosity so that $\alpha_{sat}\!=\!0$.

The left panel of fig.\,\ref{fig:saturation-amplitude-APR} shows
the static saturation amplitude for the $m=2$ r-mode of a $1.4\, M_{\odot}$
neutron star. Although this might be hard to see in certain regions
of the plot, the saturation amplitude rises steeply, within a narrow
interval, from zero at the boundary of the instability region towards
its interior. At large frequencies it reaches a plateau value that
is nearly independent of the temperature as predicted by the analytic
expression eq.\,(\ref{eq:analytic-saturation-amplitude}). As described
by the latter expression, inside the instability region the saturation
amplitude rises strongly with decreasing frequency and since the instability
region shrinks in width and eventually ends at low frequencies where
the amplitude vanishes and the mode is damped, it features a peak-like
structure. The maximum static saturation amplitude reached at the
peak is in this case unphysically large whereas the plateau value
at the Kepler frequency is still roughly $3.5$. The suprathermal
bulk viscosity can therefore in principle saturate r-modes within
the lower part of the instability region. However, in the present
case, where only the damping of the core is taken into account, the
static saturation amplitudes are at the limit where a standard r-mode
analysis is valid. Moreover these amplitudes are so far larger than
those of alternative saturation mechanisms \cite{Bondarescu:2007jw,Arras:2002dw,Gressman:2002zy,Lin:2004wx}.
It is interesting to mention once more, though, that the extreme radial
dependence of the r-mode profile eq. (\ref{eq:r-mode-profile}) strongly
weights the outer regions of the star due to power law dependences
of the inverse bulk viscosity damping time eq. (\ref{eq:bulk-damping-time})
with exponents $O\left(30\right)$ for neutron and hybrid stars which
is further enhanced by the density dependence of the inverse speed
of sound. The contribution of the crust could thereby be decisive
to obtain a realistic estimate of the impact of the non-linear viscosity.
In this context it is also important that a similar enhancement of
the bulk viscosity has been found for superfluid matter \cite{Alford:2011df}.
As noted in \cite{Alford:2010fd} there is a second instability region
at high temperatures above $10^{11}$ K and as argued above the suprathermal
bulk viscosity cannot saturate the r-mode in this high temperature
regime. It is an interesting question if the r-mode can become large
during this initial part of a star's evolution, and if so whether
the r-mode is saturated at sufficiently small values by other non-linear
mechanism or if the r-mode growth is not stopped before it reaches
the regime where the structural stability of the star is at stake.
In the latter case this instability phase might extend the violent
supernova stage and actively shape the remnant by additional mass
shedding and thereby determine its initial size and angular momentum.
We will discuss these points further in the conclusion.

The saturation amplitude for the $1.4\, M_{\odot}$ strange star  is
shown on the left panel of fig.\,\ref{fig:saturation-amplitude-Bag}.
Since the maximum of the stability window is above the Kepler frequency
there are in this plot two separate parts of the instability region.
As predicted by eq.\,(\ref{eq:analytic-saturation-amplitude}) the
plateau value of the saturation amplitude in the lower part is approximately
temperature and frequency independent. Strikingly it is of similar
size as the saturation amplitude for the $1.4\, M_{\odot}$ neutron
star at its Kepler frequency. The high temperature part of the instability
region where the viscosity again cannot saturate the r-mode extends
in this case down to lower temperatures than for neutron stars.

The left panel of fig. \ref{fig:saturation-amplitude-Hybrid} shows
the corresponding plot for the $1.4\, M_{\odot}$ hybrid star. As
found previously in \cite{Alford:2010fd} the instability region has
here three parts that are separated by two stability windows arising
from the resonant behavior of the bulk viscosities in the different
shells. Due to their parametrically different temperature dependence
the bulk viscosity of the quark shell dominates at low temperature,
whereas the bulk viscosity of the hadronic shell dominates at high
temperatures. Correspondingly the saturation amplitude in the low
temperature part of the instability region shows the qualitative behavior
found for strange matter whereas the intermediate temperature part
shows the qualitative behavior found for hadronic matter. Since the
region where the peak in fig.\,\ref{fig:saturation-amplitude-APR}
is located is {}``cut out'' by the stability window, the remaining
peak of the hadronic intermediate part of the instability region in
fig.\,\ref{fig:saturation-amplitude-Hybrid} reaches only a much
lower amplitude.

The result for the heavy $2.0\, M_{\odot}$ neutron star  is given
on the right panel of fig.\,\ref{fig:saturation-amplitude-APR}.
As had been found previously in \cite{Alford:2010fd} the instability
region is larger for such heavy stars. The figure shows in addition
that saturation occurs at a somewhat higher amplitude. The $2.0\, M_{\odot}$
strange star  is given on the right panel of fig.\,\ref{fig:saturation-amplitude-Bag}.
In this case the maximum of the stability window is below the Kepler
frequency. Similar to the high temperature behavior discussed before,
the r-mode cannot be damped by viscous effects above this maximum.
It is interesting to recall from \cite{Alford:2010fd} that in the
case of quark matter an approximate analytic expression for the location
of the maximum of the stability window exists

\begin{align}
\Omega_{max}^{\left(SS\right)} & \approx\frac{0.434m_{s}^{4/3}R^{1/3}}{\left(1-c\right)^{1/3}G^{1/3}M^{2/3}}\\
T_{max}^{\left(SS\right)} & \approx\frac{0.210\left(1-c\right)^{1/3}m_{s}^{2/3}R^{1/6}}{\hat{\Gamma}^{1/2}G^{1/6}\mu_{q}^{3/2}M^{1/3}}\end{align}
where $\hat{\Gamma}\equiv\tilde{\Gamma}/\mu_{q}^{5}$. This shows
that in addition to a large star mass, a small effective strange quark
mass in the quark matter equation of state eq.\,(\ref{eq:quark-eos-model})
increases the total instability region both at high frequency and
high temperature. In contrast, for the heavy $2.0\, M_{\odot}$ hybrid
star shown on the right panel of fig.\,\ref{fig:saturation-amplitude-Hybrid}
such a total instability region does not arise since although the
quark core cannot saturate the r-mode, the hadronic shell alone still
provides sufficient damping to do so. In summary r-modes in massive
stars are more unstable than in light stars since both their small
amplitude instability regions are larger and they are less efficiently
saturated by the large amplitude enhancement of the bulk viscosity.

The left panel of fig.\,\ref{fig:saturation-amplitude-APR-dir} shows
the static saturation amplitude for a neutron star with $2.21\, M_{\odot}$
which is the maximum mass allowed by the APR equation of state. In
this case direct Urca reactions are possible in a small inner core
region of mass $0.85\, M_{\odot}$. As had already been observed in
\cite{Alford:2010fd}, direct Urca reactions only slightly alter the
instability boundary by a small notch at its right hand side. Since
suprathermal damping from outer layers dominates due to the strong
radial dependence of the r-mode the static saturation amplitude is
likewise only slightly reduced by the small direct Urca core. However,
because the size of the inner direct Urca core depends strongly on
the equation of state and there are equations of state where the direct
Urca core is considerably larger, we show on the right panel of fig.\,\ref{fig:saturation-amplitude-APR-dir}
for comparison the (unphysical) case that the direct Urca reactions
are artificially switched on in the entire core. This represents an
upper limit for the possible effect of direct Urca reactions and shows
that in this extreme case the static saturation amplitude at large
frequency is reduced and the increase towards lower frequencies is
considerably weakened according to the $1/\Omega$-behavior predicted
by eq. (\ref{eq:analytic-saturation-amplitude}).

In contrast to the previous results that evaluated the damping time
eq.\,(\ref{eq:bulk-damping-time}) numerically the top, left panel
of fig.\ref{fig:instability-regions-multipoles} employs the approximate
analytic expression eq.\,(\ref{eq:linear-bulk-vicosity-time}) for
the $m=2$ mode of the $1.4\, M_{\odot}$ neutron star. Comparing
it to the numerical result in fig.\,\ref{fig:saturation-amplitude-APR}
shows that the corrections are very small and because the maximum
of the bulk viscosity of hadronic matter with modified Urca reactions
is reached only for large amplitudes, eq.\,(\ref{eq:linear-bulk-vicosity-time})
provides a very good approximation in this case. In contrast, the
use of the linear approximation which neglects the large amplitude
decrease of the bulk viscosity, strongly overestimates the damping
for the case of strange stars and misses the previously discussed
total instability region at high frequency in fig.\,\ref{fig:saturation-amplitude-Bag}.

Fig.\,\ref{fig:instability-regions-multipoles} also shows the saturation
amplitudes of different multipole r-modes, given for the first four
multipoles $m=2$ to $5$ of the $1.4\, M_{\odot}$ neutron star .
The higher multipoles saturate at lower amplitudes than the $m=2$
and therefore the use of the linear approximate is well justified
in this case. Interestingly, although the right segments of the lower
part of the instability boundary of these higher order r-modes had
recently been shown to be very similar to that of the fundamental
$m=2$ mode \cite{Alford:2010fd}, fig.\,\ref{fig:instability-regions-multipoles}
shows that although the peak value of the saturation amplitude of
these modes decreases, the value at the Kepler frequency stays nearly
constant. Therefore, these higher multipoles could be relevant for
the spin-down evolution since the spin-down torque due to gravitational
wave emission depends strongly on the amplitude \cite{Owen:1998xg};
for sufficiently small amplitude modes this dependence is quadratic.
So if the suprathermal damping is responsible for the r-mode saturation,
the restriction to the lowest order mode, that had been employed in
all present analyses, should present only a first approximation.

\begin{figure*}
\begin{minipage}[t]{0.5\textwidth}%
\includegraphics[scale=0.85]{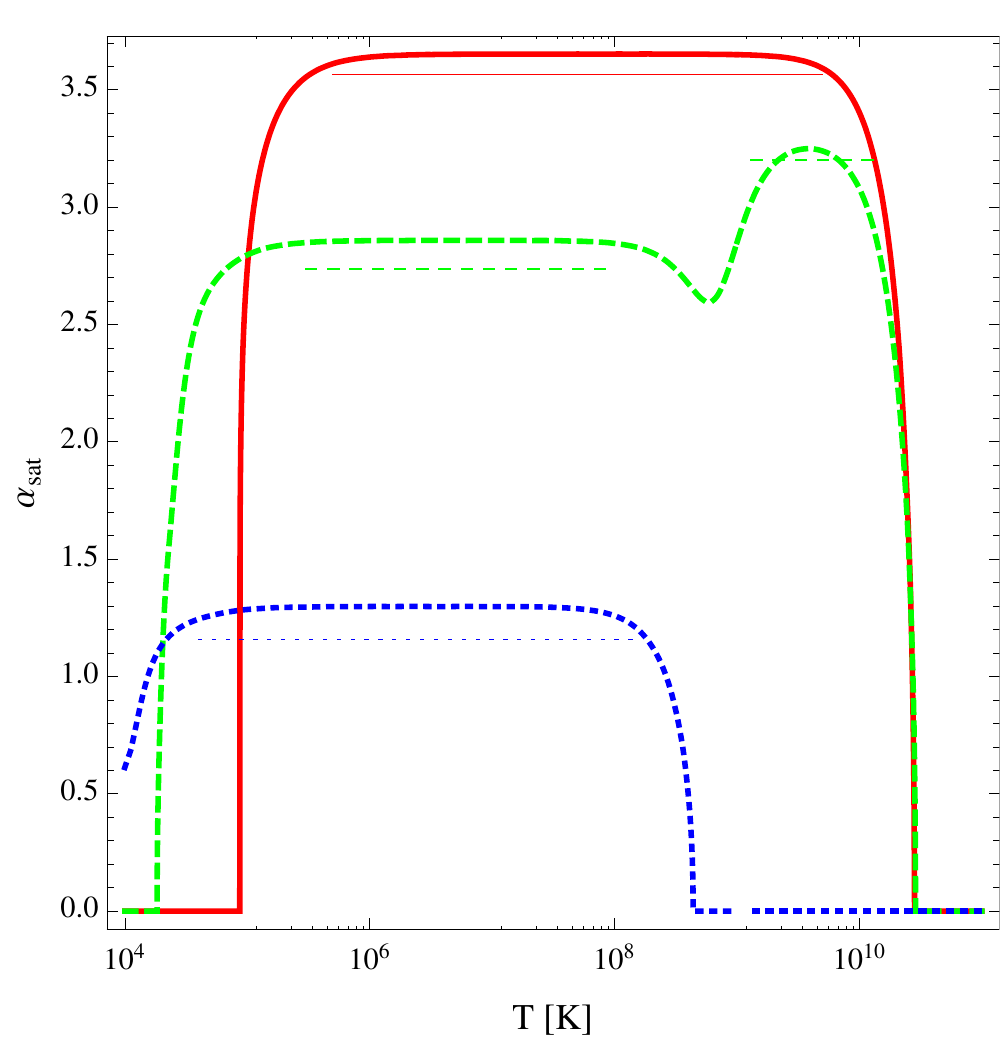}%
\end{minipage}%
\begin{minipage}[t]{0.5\textwidth}%
\includegraphics[scale=0.85]{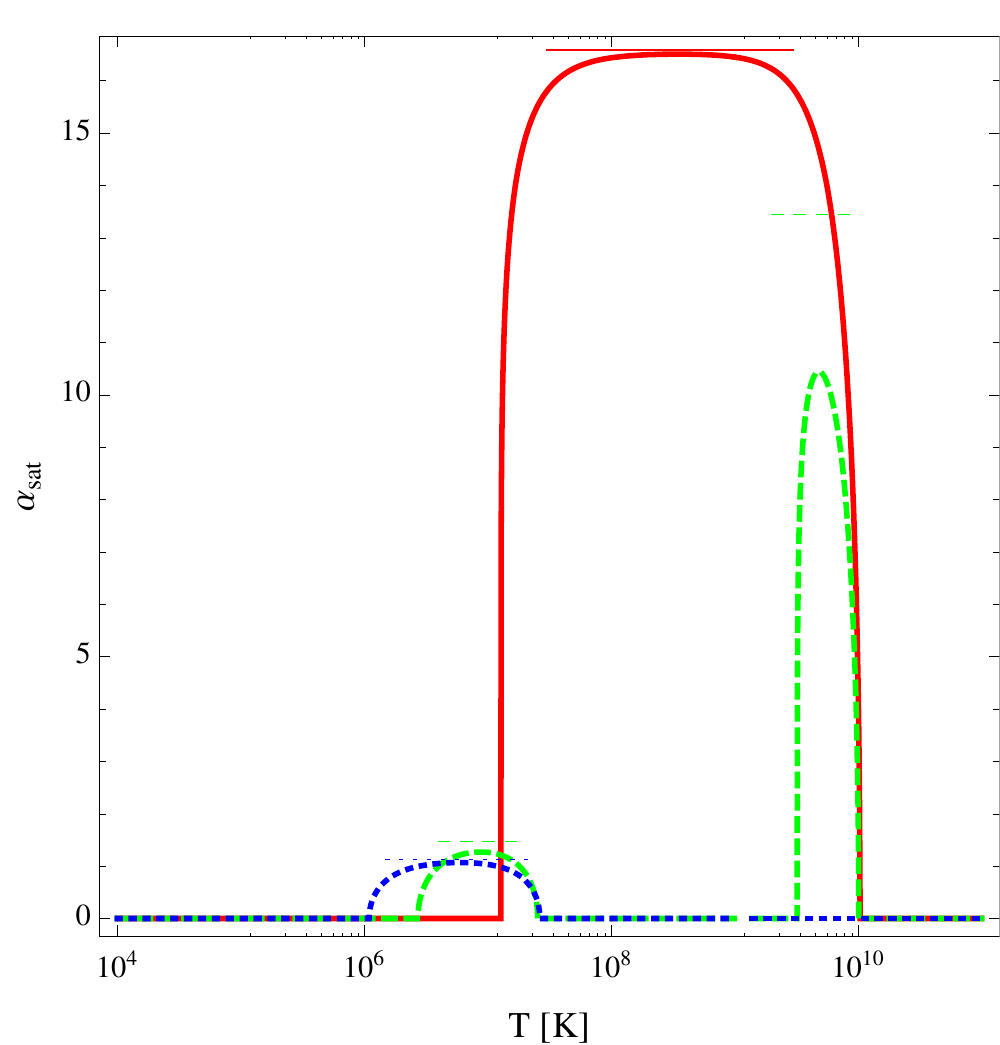}%
\end{minipage}

\caption{\label{fig:sat-amps-ms}Comparison of the saturation amplitudes for
the different $1.4\, M_{\odot}$ stars. \emph{Left panel:} Stars spinning
with a period of $1$ ms. \emph{Right panel:} Same for stars rotating
with a period of $4$ ms. Shown are the considered neutron star (solid),
the hybrid star (dashed) and the strange star (dotted). The thick
curves present the numerical results and the thin horizontal segments
denote the analytic values obtained from eq.\,(\ref{eq:analytic-saturation-amplitude}). }

\end{figure*}

The saturation amplitudes of the different $1.4\, M_{\odot}$ stars
are finally compared with each other and the analytic approximation
eq.\,(\ref{eq:linear-bulk-vicosity-time}) in fig.\,\ref{fig:sat-amps-ms}.
Surprisingly, all stars feature saturation amplitudes of the same
order of magnitude for millisecond pulsars, despite their very different
microscopic and structural aspects. As noted before, for larger oscillation
periods hadronic stars and to some extent also hybrid stars feature
considerably larger saturation amplitudes than strange stars due to
the parametrically different frequency dependence, see eq.\,(\ref{eq:analytic-saturation-amplitude}).
The analytic approximation yields in most cases a reasonable approximation
to the full results with errors below the $10\%$ level. In general
the analytic result overestimates the actual amplitude since it only
describes the result far away from the boundaries and boundary effects
play a role. In contrast at large frequencies the analytic approximation
underestimates the saturation amplitude since the considered frequencies
are already close to the critical values, where the saturation amplitude
diverges, see fig.\,(\ref{fig:saturation-amplitude-Bag}). Nevertheless,
the analytic approximation provides an important and reliable estimate
for the order of magnitude of the static saturation amplitude which,
as will be discussed in more detail below, provides an upper limit
for saturation amplitudes taken in dynamical star evolutions.

\section{Conclusions}

Using the recent general results for the bulk viscosity that include
its non-linear behavior at large amplitudes we have derived expressions
for the r-mode damping time that show that in the regime below the
resonant temperature of the bulk viscosity, large amplitude r-modes
are damped on considerably shorter time scales than low amplitude
oscillations. In contrast, the universal maximum of the bulk viscosity
found in \cite{Alford:2010gw} implies that at very high temperatures
and frequencies r-modes cannot be damped at all by viscous effects
since there is no enhancement in the suprathermal limit. We find that
for most stars considered in this work the corresponding critical
frequency is above the Kepler frequency. On the other hand the r-modes
of all considered stars are unstable at temperatures that are expected
to be present when a proto-neutron star is created. At lower temperatures
our results lead to an extension of the concept of the instability
region of an r-mode since the latter is only initially unstable at
small amplitudes but the suprathermal viscous damping can saturate
the r-mode growth at finite amplitudes. We find that well within the
instability region the static saturation amplitude $\alpha_{sat}$
defined in the text is temperature independent and takes values $O\!\left(1\right)$
at milli-second frequencies for all considered stars. This is incidentally
the order of magnitude that had been assumed in early r-mode analyses
\cite{Owen:1998xg}. Yet, the static values obtained here represent
only an upper limit for the actual amplitude reached in the dynamic
evolution. Our numeric results are confirmed by approximate analytic
expressions which reveal the dependence of these results on the various
underlying parameters. We also studied higher multipoles and find
that although the first few multipoles have instability regions that
are sizable, they feature similar saturation amplitudes as the fundamental
$m\!=\!2$ mode for millisecond pulsars and could thereby be relevant.

It is interesting to compare our saturation mechanism and the obtained
results for the saturation amplitudes with previously proposed mechanisms.
In general when there are different competing saturation mechanisms,
the one with the smallest saturation amplitude should dominate and
effectively saturate the mode. Explicit numerical analyses of the
general relativistic hydrodynamical equations \cite{Lindblom:2000az,Gressman:2002zy,Lin:2004wx}
would present the ideal way to study the saturation and star evolution.
Whereas some of these studies find saturation only at large amplitudes,
in others the r-mode can be completely destroyed by the decay into
daughter modes once it exceeds amplitudes $O\left(10^{-2}\right)$
\cite{Lin:2004wx}, see also \cite{Brink:2004bg}. However, the numerical
complexity limits these analyses so far to unphysically large values
of the radiation reaction force that are orders of magnitude above
the physical value and it is not clear to what extent the obtained
results can be extrapolated to the physical case.  Another proposed
saturation mechanism relies on the non-linear coupling of different
oscillation modes \cite{Arras:2002dw,Bondarescu:2007jw,Bondarescu:2008qx,Brink:2004bg}.
These analyses find that this mechanism could saturate r-modes at
amplitudes as low as $O\left(10^{-5}\right)$. Due to the considerable
difficulties of a complete description of such a mode coupling mechanism,
these analyses have to rely on model systems of generic coupled oscillators
without a detailed connection to the complicated coupling of collective
star oscillations. In summary, within the present approximation to
neglect the neutron star crust, competing mechanisms will very likely
dominate and saturate the r-mode at lower values than the suprathermal
enhancement of the viscosity. However, these mechanisms still involve
simplifications and uncertainties. Our novel saturation mechanism,
in contrast, relies on standard viscous effects and microscopic physics
that is quantitatively well understood.

Let us now discuss the implications of our results for the spin-down
of compact stars. In the supernova formation process where a much
larger star contracts to a very compact object that takes over the
angular momentum it seems plausible that fast rotating proto-neutron
stars could be formed which spin with frequencies close to the Kepler
limit. According to our results r-mode oscillations are unstable in
this initial hot stage $T\!\gtrsim\!10^{10}K$ and cannot be saturated
by viscous effects for all considered forms of dense matter. Generically,
the cooling is very fast in this regime so that the evolution could
leave this instability region before large amplitude r-modes develop
or spin down the star. The star will then cool until it reaches the
lower instability zone and the r-mode develops. According to fig.\,\ref{fig:sat-amps-ms}
in this regime the r-mode can be saturated by viscous damping. For
strange stars such a saturation does not seem to be required at all
since the instability region is in this case located at comparably
low temperatures \cite{Madsen:1998qb} where the cooling becomes slow
and the star either quickly spins down \cite{Andersson:2001ev}, or
when reheating effects are considered it reheats again \cite{Drago:2007iy},
and leaves the instability region before the amplitude becomes large.
In this case the evolution wiggles around the instability line thereby
spinning down the star, but this can take billions of years due to
the strong reheating. 

In contrast in the case of neutron and hybrid stars without strangeness,
the instability region is reached at large temperatures where cooling
is still fast and reheating effects are moderate, so that the evolution
quickly penetrates the instability region and a saturation mechanism
is indeed required to stop the r-mode growth \cite{Owen:1998xg}.
Since the static saturation amplitude increases continuously at the
boundary of the instability region the discussed \emph{static} value
does not have to be reached but a \emph{dynamic} equilibrium could
be established at a lower saturation amplitude that is reached once
the r-mode is sufficiently large that the spindown becomes efficient.
Due to this the viscous saturation could dominate competing saturation
mechanisms. Once the r-mode is saturated, the question is which one
of two competing processes, cooling or spin-down, is faster. Since
the cooling is slowing down at lower temperatures it is likely that
the spin-down wins and the evolution leaves the instability region
near its lower boundary. In this case no young compact stars with
frequencies larger than a tenth of the Kepler frequency would be possible
which is in good agreement with observations. An answer to the above
questions requires a detailed study of the combined spin-down and
cooling evolution of the star which will be presented elsewhere.

Strikingly our results suggest even another possibility for the spindown
of young stars that would be even faster and more violent. The core
bounce during the supernova process should excite rather large amplitude
oscillation modes in the forming compact core. Since r-modes are unstable
in this regime \cite{Alford:2010fd} these will grow further. Because
of the initial high temperatures, neutrinos are trapped inside the
proto-neutron stars for roughly a minute \cite{Prakash:1996xs}. Since
a neutron star crust, that could provide an efficient damping mechanism
\cite{Bildsten:2000ApJ...529L..33B}, is not formed at this point
and as our results show viscous effects cannot stop the r-mode growth,
the amplitude could indeed become large if other non-linear saturation
mechanisms likewise cannot operate efficiently in this turbulent environment.
In this case the loss of angular momentum could proceed not by gravitational
wave emission but by actual mass shedding and thereby effectively
as an extension of the supernova explosion that is driven by r-modes.
Since such a violent spindown should be fast the star could end up
at the lower boundary of the high temperature instability region before
the star becomes transparent to neutrinos and the cooling process
starts. Clearly, in this initial stage, which cannot rigorously be
separated from the aftermath of the supernova explosion, the dynamics
is highly non-linear and our simple r-mode analysis might not directly
apply. Whether such a mechanism is feasible will therefore require
further study, but this mechanism would naturally explain the observed
absence of fast, young pulsars independent of their internal composition
and it is striking that the frequencies of the high temperature instability
boundary also seem to agree well with fastest pulsars that are young
enough that they cannot be spun up by accretion \cite{Manchester:2004bp}.

Finally, r-modes should also be relevant for old accreting stars in
binary systems that are spun up and could enter the instability region
at low temperatures from below \cite{Reisenegger:2003cq}. As discussed
in \cite{Alford:2010fd}, strange and hybrid stars feature stability
windows at low temperatures where the r-mode is absent, so that such
stars could accelerate to frequencies close to the Kepler frequency.
In contrast for neutron stars there is no stability window at low
temperatures so that an accreting star would enter the unstable regime
already at low frequencies. Recall that the saturation amplitude of
neutron stars due to bulk viscosity has a characteristic form with
a pronounced peak close to the minimum of the instability region.
In case the r-mode is saturated by suprathermal bulk viscosity, the
steep rise of the amplitude close to the maximum should spin down
the star quickly and so that it cannot penetrate deep into the instability
region. This means that such stars should cluster close to the boundary
which might be a signature once more observational data for the temperature
of compact stars becomes available.

\begin{acknowledgments}
We thank Nils Andersson, Greg Comer, Brynmor Haskell, Prashant Jaikumar,
Andreas Reisenegger, Andrew Steiner and Ira Wasserman for helpful
discussions. This research was supported in part by the Offices of
Nuclear Physics and High Energy Physics of the U.S. Department of
Energy under contracts \#DE-FG02-91ER40628, \#DE-FG02-05ER41375. 
\end{acknowledgments}
\appendix

\section{R-mode and amplitude conventions\label{sec:R-mode-profile}}

In this appendix we review a few standard expressions for r-modes
and discuss different conventions used in the literature. The form
of the r-mode oscillation is most conveniently derived \cite{Ipser:1989zz}
from the underlying equations that determine the fluctuation of the
potential $\delta U=\delta h+\delta\Phi$, where $h$ is the enthalpy
and $\Phi$ the gravitational potential, since then the hydrodynamic
Euler equation for the harmonic, cylindrically symmetric perturbation
reduces from a differential to an ordinary linear equation and can
be solved analytically by matrix inversion. The expression for $\delta U$
reads to leading order in a slow rotation expansion

\begin{equation}
\delta U\approx\sqrt{\frac{m}{\pi\left(m+1\right)^{3}\left(2m+1\right)!}}\alpha R^{2}\Omega^{2}\left(\frac{r}{R}\right)^{m+1}P_{m+1}^{m}\left(\cos\theta\right)\mathrm{e}^{im\phi}\label{eq:potential-fluctuation}\end{equation}
The velocity fluctuation is obtained from $\delta U$ by application
of a differential operator \cite{Ipser:1989zz} and yields to leading
order in a slow rotation expansion \cite{Lindblom:1998wf} 

\[
\delta\vec{v}=\alpha R\Omega\left(\frac{r}{R}\right)^{m}\vec{Y}_{mm}^{B}\mathrm{e}^{i\omega t}\]
This expression provides the definition of $\alpha$ within the convention
used here. In spherical coordinates this expression yields the explicit
form

\begin{align*}
\delta\vec{v}= & \frac{\left(-1\right)^{m}}{2^{m}m!}\sqrt{\frac{m\left(2m+1\right)\left(2m\right)!}{4\pi\left(m+1\right)}}\alpha R\Omega\\
 & \quad\cdot\left(\frac{r}{R}\right)^{m}\left(\sin\theta\right)^{m-1}\mathrm{e}^{i\left(m\phi+\omega t\right)}\left(-i\hat{\theta}+\cos\theta\hat{\phi}\right)\end{align*}
where $\hat{\theta}$ and $\hat{\phi}$ are unit vectors in polar
and azimuthal direction. With this definition and for $\alpha=1$
the maximum value, taken at the equator and in direction of $\hat{\theta}$,
is roughly $\delta\vec{v}/\vec{v}\approx0.3$, so that the approximation
breaks down for $\alpha\gg1$ since mass shedding will occur for fast
spinning stars. The corresponding maximum density fluctuation $\delta n/\bar{n}$
obtained from eq. (\ref{eq:r-mode-profile}) is more than an order
of magnitude smaller. 

In contrast in \cite{Lindblom:1999yk} an alternative convention $\alpha^{\prime}$
of the amplitude was introduced that is related to the above $\alpha$
by

\[
\alpha=\sqrt{\frac{\pi\left(m+1\right)^{3}\left(2m+1\right)!}{m}}\alpha^{\prime}\]
i.e. defined by eq. (\ref{eq:potential-fluctuation}) without the
algebraic prefactor.

\section{Approximate result for strange quark matter\label{sec:strange-matter-approximation}}

Whereas the bulk viscosity of semi-leptonic processes in hadronic
and quark matter require in general a numeric solution, for the dominant
contribution from non-leptonic processes in strange quark matter an
approximate solution valid in both subthermal and suprathermal regimes
is possible. The approximate analytic result for the bulk viscosity
in the suprathermal regime obtained from a Fourier analysis, is given
by \cite{Alford:2010gw}

\begin{equation}
\zeta^{>}\approx\frac{2}{3\sqrt{3}}\frac{C^{2}}{B\omega}\, h\biggl(\frac{9\sqrt{3}\chi}{8}\frac{\tilde{\Gamma}BC^{2}T^{\delta-2}}{\omega}\Bigl(\frac{\Delta\! n_{*}}{\bar{n}_{*}}^{2}\Bigr)\biggr)\label{eq:supra-viscosity}\end{equation}
where $\chi\!\equiv\!\chi_{1}$ and

\begin{equation}
h(z)=\frac{9}{4z}\left(\left(\sqrt{z^{2}\!+\!1}\!-\! z\right)^{\frac{2}{3}}\!+\!\left(\sqrt{z^{2}\!+\!1}\!+\! z\right)^{\frac{2}{3}}\!-\!2\right)\label{eq:h-function}\end{equation}
A very good parameterization valid in both the sub- and suprathermal
regime is given by \cite{Alford:2010gw}

\begin{equation}
\zeta_{par}\approx\zeta^{<}+\theta(T_{max}-T)\frac{\zeta_{max}-\zeta^{<}}{\zeta_{max}}\zeta^{>}\label{eq:viscosity-parametrization}\end{equation}
Since the functional form of the suprathermal viscosity eq.\,(\ref{eq:h-function})
is still complicated and does not allow to perform the necessary subsequent
integrations to obtain the r-mode damping time in an analytic form,
we give an approximate analytic result that is valid up to the maximum
of the bulk viscosity. To this end we perform a global polynomial
interpolation to the function $h\!(z)$ in the interval $\left[0,z_{max}\right]$.
In order to appropriately describe the low amplitude behavior and
the qualitative form below the maximum requires at least a quartic
polynomial which is then uniquely determined as

\begin{equation}
h_{pol}(z)=z-\frac{1}{2\sqrt{3}}z^{2}+\frac{1}{27}z^{3}-\frac{1}{324\sqrt{3}}z^{4}\label{eq:polynomial-approximation}\end{equation}
The leading linear term in eq.\,(\ref{eq:polynomial-approximation})
reproduces the approximate intermediate linear result given by Madsen
\cite{Madsen:1992sx}, whereas the other terms ensure the proper large
amplitude saturation. The analytic form and the polynomial approximation
agree in the relevant region below the maximum point-wise on the
5\% level and the corresponding integrals required for the damping
time to even better accuracy.

The density in a strange star is nearly constant and so the density
dependent quantities can be approximated by their value at the radius
of the star, denoted by the suffix $R$. Performing the integration
over the r-mode profile eq. (\ref{eq:r-mode-profile}), we find the
approximate analytic result for the viscous damping time

\begin{widetext}

\begin{equation}
\frac{1}{\tau_{B}^{>}}\approx\frac{16}{5103}\frac{A_{R}^{2}C_{R}^{2}\Omega^{3}R^{5}}{B_{R}M\tilde{J}}\left(\frac{\frac{3}{2}\Omega\tilde{\Gamma}_{R}B_{R}T^{\delta}}{\Omega^{2}+\frac{9}{4}\tilde{\Gamma}_{R}^{2}B_{R}^{2}T^{2\delta}}+\frac{2430}{143}\theta\bigl(T_{max}-T\bigr)\frac{(\Omega-\frac{3}{2}\tilde{\Gamma}_{R}B_{R}T^{\delta})^{2}}{\Omega^{2}+\frac{9}{4}\tilde{\Gamma}_{R}^{2}B_{R}^{2}T^{2\delta}}g\bigl(\chi\tilde{\Gamma}_{R}A_{R}^{2}B_{R}C_{R}^{2}R^{4}\Omega^{3}T^{\delta-2}\alpha^{2}\bigr)\right)\label{eq:homogenous-quark-damping-time}\end{equation}
\end{widetext}where $g$ is the polynomial

\[
g(x)=x-\frac{151875}{9044}x^{2}+\frac{1063125}{7429}x^{3}-\frac{290631796875}{587723048}x^{4}\]
\bibliographystyle{JHEP_MGA}
\bibliography{cs}

\end{document}